\documentclass[aps,prl,twocolumn,superscriptaddress,preprintnumbers,floatfix,10pt,nofootinbib]{revtex4-1}
\pdfoutput=1
\usepackage[bookmarks=false]{hyperref}
\usepackage{amsmath}
\usepackage{graphicx}
\usepackage[small]{subfigure}
\usepackage{xcolor}

\newcommand{\refcite}[1]{Ref.~\cite{#1}}
\newcommand{\refscite}[1]{Refs.~\cite{#1}}
\newcommand{\eq}[1]{Eq.~\eqref{eq:#1}}
\newcommand{\eqs}[2]{Eqs.~\eqref{eq:#1} and \eqref{eq:#2}}

\newcommand{\fig}[1]{Fig.~\ref{fig:#1}}
\newcommand{\figs}[2]{Figs.~\ref{fig:#1} and \ref{fig:#2}}

\newcommand{\tab}[1]{Table~\ref{tab:#1}}

\newcommand{\nn}{\nonumber}

\newcommand{\abs}[1]{\lvert#1\rvert}

\newcommand{\ord}[1]{\mathcal{O}(#1)}

\newcommand{\ORd}[1]{\mathcal{O}\Bigl(#1\Bigr)}
\newcommand{\ordsq}[1]{\mathcal{O}[#1]}

\newcommand{\df}{\mathrm{d}}
\newcommand{\img}{\mathrm{i}}

\let\oldvec\vec
\renewcommand*\vec[1]{\oldvec{\kern0pt #1}}

\renewcommand*{\tt}{\tilde{t}}
\newcommand{\tB}{\tilde{B}}
\newcommand{\tI}{\tilde{I}}
\newcommand{\hS}{\hat{S}}
\newcommand{\hsigma}{\hat{\sigma}}
\newcommand{\heta}{\hat{\eta}}
\newcommand{\bn}{{\bar{n}}}
\newcommand{\bq}{{\bar{q}}}
\newcommand{\bz}{{\bar{z}}}
\newcommand{\cI}{\mathcal{I}}
\newcommand{\cL}{\mathcal{L}}

\newcommand{\tcI}{\tilde{\mathcal{I}}}
\newcommand{\hcI}{\hat{\mathcal{I}}}

\newcommand{\lqcd}{\lambda_\mathrm{QCD}}
\newcommand{\LQCD}{\Lambda_\mathrm{QCD}}

\newcommand{\umax}{u_\mathrm{max}}

\newcommand{\cusp}{\mathrm{cusp}}

\newcommand{\DY}{\mathrm{DY}}
\newcommand{\gen}{\mathrm{gen}}
\newcommand{\ggH}{{ggH}}
\newcommand{\high}{\mathrm{high}}

\newcommand{\low}{\mathrm{low}}

\newcommand{\soft}{\mathrm{soft}}
\newcommand{\thr}{\mathrm{thr}}

\newcommand{\as}{\alpha_s}
\newcommand{\Ecm}{E_\mathrm{cm}}
\newcommand{\Ymax}{Y_\mathrm{max}}

\allowdisplaybreaks[2]

\begin{document}


\preprint{\vbox{\hbox{DESY 19-135}\hbox{NIKHEF 2019-021}}}

\title{Generalized Threshold Factorization with Full Collinear Dynamics}

\author{Gillian Lustermans}%
\email{g.h.h.lustermans@uva.nl}%
\affiliation{Institute for Theoretical Physics Amsterdam and Delta Institute for Theoretical Physics, University of Amsterdam, Science Park 904, 1098 XH Amsterdam, The Netherlands}%
\affiliation{Nikhef, Theory Group, Science Park 105, 1098 XG, Amsterdam, The Netherlands}%

\author{Johannes K.~L.~Michel}%
\email{johannes.michel@desy.de}%
\affiliation{Theory Group, Deutsches Elektronen-Synchrotron (DESY), D-22607 Hamburg, Germany}%

\author{Frank J.~Tackmann,}%
\email{frank.tackmann@desy.de}%
\affiliation{Theory Group, Deutsches Elektronen-Synchrotron (DESY), D-22607 Hamburg, Germany}%

\date{August 2, 2019}

\begin{abstract}

Soft threshold factorization has been used extensively to study hadronic
collisions. It is derived in the limit where the momentum fractions $x_{a,b}$ of
both incoming partons approach $x_{a,b}\to 1$. We present a generalized
threshold factorization theorem for color-singlet processes, which holds in the
weaker limit of only $x_a \to 1$ for generic $x_b$ (or vice versa),
corresponding to the limit of large rapidity but generic invariant mass of the
produced color singlet.
It encodes the complete soft and/or collinear singular structure in the partonic
momentum fractions to all orders in perturbation theory, including in particular
flavor-nondiagonal partonic channels at leading power. It provides a more
powerful approximation than the classic soft threshold limit, capturing a much
larger set of contributions.
We demonstrate this explicitly for the $Z$ and Higgs rapidity spectrum to NNLO, and
we use it to predict a nontrivial set of its N$^3$LO contributions.
Our factorization theorem provides the relevant resummation of large-$x$
logarithms in the rapidity spectrum required for resummation-improved
PDF fits.
One of our factorization ingredients is a new beam function closely related to
the $N$-jettiness beam function. As a byproduct, we identify the correct soft
threshold factorization for rapidity spectra among the differing results in the
literature.

\end{abstract}

\maketitle

\section{Introduction}

Color-singlet processes play a central role in the LHC physics program.
The $pp\to Z,W$ Drell-Yan processes are precision benchmarks, providing
determinations of electroweak parameters and important inputs
for fits of parton distribution functions (PDFs). Higgs and diboson processes
provide strong sensitivity to possible contributions beyond the Standard Model.

We consider the production of a generic color-singlet final state $L$ together
with hadronic radiation $X$,
\begin{align}
p(P_a^\mu) + p (P_b^\mu) \to L(q^\mu) + X(P_X^\mu)
\,,\end{align}
at hadronic center-of-mass energy $\Ecm^2 = (P_a + P_b)^2$. The key observables
characterizing $L$ are its total invariant mass $Q \equiv \sqrt{q^2}$,
rapidity $Y$, and transverse momentum $q_T \equiv \abs{\vec q_T}$.
We define the momentum fractions
\begin{align} \label{eq:def_xab}
x_a = \frac{Q}{\Ecm}e^{+Y}
\,, \quad
x_b = \frac{Q}{\Ecm}e^{-Y}
\,, \quad
\tau = x_a x_b
\,,\end{align}
which are equivalent to $Q$ and $Y$.
The cross section differential in $x_{a,b}$ is given by~\cite{Bodwin:1984hc, Collins:1985ue, Collins:1988ig}
\begin{align} \label{eq:collinear_factorization}
\frac{\df \sigma}{\df x_a \df x_b}
&= \int \! \frac{\df z_a}{z_a} \frac{\df z_b}{z_b} \, \hsigma_{ij}(z_a,z_b) \,
f_i\Bigl(\frac{x_a}{z_a}\Bigr) \, f_j\Bigl(\frac{x_b}{z_b}\Bigr)
\,,\end{align}
where $\hsigma_{ij}(z_a, z_b)$ denotes the perturbatively calculable partonic cross section and
$f_{i,j}(x)$ are the standard PDFs. We always implicitly sum over
parton indices $i, j$, and keep the dependence on renormalization scales $\mu$ implicit.

In the soft threshold limit $\tau \to 1$, which implies that both $x_{a,b}\to 1$,
\eq{collinear_factorization} factorizes further~\cite{Sterman:1986aj, Catani:1989ne, Ravindran:2006bu, Westmark:2017uig, Banerjee:2017cfc, Banerjee:2018vvb},
\begin{align} \label{eq:factorization_soft}
\frac{\df \sigma}{\df x_a \df x_b}
&= H_{ij}(Q^2) \int \! \df k^- \, \df k^+ S(k^-, k^+)
\\\nn & \quad \times
f^\thr_i\Bigl[x_a\Bigl(1 + \frac{k^-}{Q e^{+Y}}\Bigr)\Bigr]
f^\thr_j\Bigl[x_b\Bigl(1 + \frac{k^+}{Q e^{-Y}}\Bigr)\Bigr]
.\end{align}
In this limit, the hadronic final state is forced to be soft, and is described
by the soft function $S$, which encodes soft-gluon emissions
from the colliding hard partons.
Furthermore, only the hard Born processes, e.g.\ $q\bar q \to Z$ or $gg\to H$,
contribute. They are encoded in the hard function $H_{ij}$, including the
Born-like virtual corrections.
Any nondiagonal partonic channels like $qg \to Lq$ vanish for $\tau \to 1$.
The threshold PDF $f^\thr_i(x)$ encodes the extraction of parton $i$ from the proton for $x \to 1$.
At partonic level, \eq{factorization_soft} implies that for $z \equiv z_a z_b\to 1$,
which requires both $z_{a,b}\to 1$,
\begin{align} \label{eq:factorization_soft_partonic}
\hsigma_{ij}(z_a, z_b)
&= H_{ij}\, \hS(z_a, z_b)
\,,\end{align}
up to power corrections in $1-z$.

While taking $\tau\to 1$ forces $z\to 1$, even for typical LHC values of
$\tau \ll 1$, the $z\sim 1$ region often numerically dominates the cross section.
Soft threshold factorization has thus been widely used for decades.
It enables the all-order resummation of the leading terms in $1-z$,
see e.g.~\refscite{Sterman:1986aj, Catani:1989ne, Ravindran:2006bu, Westmark:2017uig, Banerjee:2017cfc, Banerjee:2018vvb, Appell:1988ie, Magnea:1990qg, Korchemsky:1992xv, Contopanagos:1996nh, Catani:1996yz, Belitsky:1998tc, Moch:2005ky, Laenen:2005uz, Idilbi:2006dg, Mukherjee:2006uu, Bolzoni:2006ky, Becher:2007ty, Bonvini:2010tp, Bonvini:2015ira, Fuks:2013vua, Bonvini:2014joa, Schmidt:2015cea, H:2019dcl}.
The resummation at next-to-leading power (NLP) in $1-z$
has also received recent interest~\cite{Bonocore:2016awd, DelDuca:2017twk, Beneke:2018gvs}.
Another important use is to approximate the fixed-order cross section by expanding in
$1-z$, e.g.\ at N$^3$LO~\cite{Anastasiou:2014vaa, Ahmed:2014cla, Ahmed:2014uya,
deFlorian:2014vta, Li:2014afw, Anastasiou:2016cez, Dulat:2017prg, Dulat:2018bfe}.

In this letter, we derive the factorization that generalizes \eqs{factorization_soft}{factorization_soft_partonic} to the weaker limit where only one of $x_{a,b}$
(or $z_{a,b}$) approaches $1$ while keeping the
exact dependence on the other variable. This corresponds
to the kinematic limit $\abs{Y} \to \Ymax = \ln(\Ecm/Q)$
for generic (including small) $Q$ values.

\section{Generalized threshold factorization}

\begin{figure}
\includegraphics[width=0.75\columnwidth]{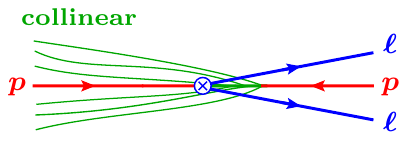}%
\caption{Illustration of Drell-Yan at large dilepton rapidity.
\label{fig:illustration_endpoint_kinematics}}
\end{figure}

We use light-cone coordinates $p^\mu \equiv (n\cdot p, \bar{n}\cdot p, p_\perp) \equiv (p^+, p^-, p_\perp)$
with respect to lightlike vectors $n^\mu \equiv (1, \hat{z})$ and $\bar{n}^\mu \equiv (1,-\hat{z})$
along the beam axis $\hat z$. We first consider the observables $q^\mp$ instead of $Q$ and $Y$,
with corresponding momentum fractions
\begin{align} \label{eq:def_txab}
x_\mp \equiv \frac{q^\mp}{P_{a,b}^\mp}
= \frac{\sqrt{Q^2 + q_T^2}}{\Ecm}\, e^{\pm Y}
\,.\end{align}
We consider the generalized threshold limit
\begin{equation}
\lqcd^2 \ll \lambda^2 \sim 1 - x_- \ll 1
\quad\text{for generic $x_+$}
\,,\end{equation}
where $\lqcd \equiv \LQCD/Q$ and $\lambda$ are power-counting parameters.
In this limit, illustrated in \fig{illustration_endpoint_kinematics}, $L$ has large $Y$
while the emissions in $X$ become collimated in the opposite direction
with typical momenta%
\begin{equation} \label{eq:pX_scaling}
p_X^\mu \sim (q^+, P_a^- - q^-, p_{X\perp}) \sim (q^+, \lambda^2 q^-, \lambda \sqrt{q^+q^-})
\,.\end{equation}

In this situation, the following factorization theorem holds to leading
power in $1 - x_-$,
\begin{align} \label{eq:factorization_collinear_endpoint_qm_qp}
\frac{\df \sigma}{\df x_- \df x_+}
&= H_{ij}(q^+ q^-) \int \! \df t  \, f_i^\thr\Bigl[x_-\Bigl(1 + \frac{t}{q^+  q^-} \Bigr)\Bigr]
\nn \\ & \quad\times
B_j(t, x_+)
\,.\end{align}
Here, $B_j(t,x)$ is the inclusive beam function that also appears in the factorization
for $N$-jettiness~\cite{Stewart:2009yx, Stewart:2010tn}. It
depends on the transverse virtuality $t$ and momentum fraction $x$ of the
colliding parton $j$.
Since $t \sim p_X^2 \sim \lambda^2 Q^2 \gg \LQCD^2$, it can be calculated perturbatively
in terms of standard PDFs as~\cite{Stewart:2009yx, Stewart:2010qs}
\begin{align} \label{eq:beam_function_matching}
B_j(t, x) = \int \! \frac{\df z}{z} \, \cI_{jk}(t, z) \, f_k\Bigl( \frac{x}{z}\Bigr)
\,.\end{align}
The matching coefficients $\cI_{jk}$ are known to
$\ord{\as^2}$~\cite{Stewart:2010qs, Berger:2010xi, Gaunt:2014xga, Gaunt:2014cfa},
with progress at $\ord{\as^3}$~\cite{Melnikov:2018jxb, Melnikov:2019pdm}.

To derive \eq{factorization_collinear_endpoint_qm_qp}, we use
Soft-Collinear Effective Theory (SCET)~\cite{Bauer:2000ew, Bauer:2000yr,
Bauer:2001ct, Bauer:2001yt, Bauer:2002nz},
which is the effective theory of QCD in the limit $\lambda \ll 1$.
The derivation proceeds in a standard fashion~\cite{Stewart:2009yx}, with
more details given in~\cite{*[{See Supplemental Material at the end of this preprint}] [{}] supplement}.
The key elements are the necessary degrees of freedom (modes) in the effective theory,
\begin{align} \label{eq:mode_setup}
p_\bn &\sim Q\, \bigl(1, \lambda^2, \lambda \bigr)
\,, \quad
P_\bn \sim Q\, (1, \lqcd^2, \lqcd)
\,, \\ \nn
P_n &\sim Q\, (\lqcd^2, 1, \lqcd)
\,, \quad
P_{s} \sim Q\, \Bigl(\frac{\lqcd^2}{\lambda^2}, \lambda^2, \lqcd\Bigr)
\,.\end{align}
The $p_\bn$ collinear modes describe the QCD final state at the scale $\lambda Q$,
which due to \eq{pX_scaling} is $\bn$-collinear.
The $P_{n,\bn}$ collinear modes describe the PDFs at the scale $\lqcd Q$.
The $P_s$ modes describe the soft interactions between the $p_\bn$
and $P_n$ modes.
Possible ultrasoft/Glauber modes $P_{us,G}\sim Q (\lqcd^2, \lqcd^2, \lqcd^{(2)})$
cancel as in \eq{collinear_factorization} because the measurement is the same
and its scale $\lambda Q\gg \lqcd Q$.

The hard function $H_{ij}$ in \eq{factorization_collinear_endpoint_qm_qp}
is the same as in \eq{factorization_soft}. It arises from matching the electroweak
current for $L$ onto the corresponding SCET current.
There are no interactions between the modes in the leading-power SCET Lagrangian,
so the complete matrix element factorizes into separate ones in each sector.
The matrix element of the combined $p_\bn$ and $P_\bn$ modes yields
$B_j(t, x)$, and their separation leads to \eq{beam_function_matching}~\cite{Stewart:2009yx, Stewart:2010qs}.
The combined matrix element of the $P_n$ and $P_s$ modes yields
$f_i^\thr$~\cite{Fleming:2012kb, Hoang:2015iva}.
The convolution structure in \eq{factorization_collinear_endpoint_qm_qp} follows
from momentum conservation~\cite{supplement}.

Next, we consider also measuring $\vec q_T$.
From \eq{pX_scaling}, it follows that generically
$q_T \sim p_{X\perp}\sim \lambda Q$, so the $q_T$ dependence is entirely described
by the $p_\bn$ modes, which yields the factorization theorem
\begin{align} \label{eq:factorization_collinear_endpoint_qm_qp_qT}
\frac{\df \sigma}{\df x_- \df x_+ \, \df \vec{q}_T}
&= H_{ij}(q^+ q^-) \int \! \df t  \, f_i^\thr \Bigl[x_-\Bigl(1 + \frac{t}{q^+ q^-} \Bigr)\Bigr]
\nn\\ &\quad\times
B_j(t, \vec{q}_T, x_+)
\,.\end{align}
Here, $B_j(t, \vec{k}_T, x)$ is the double-differential beam function~\cite{Jain:2011iu, Gaunt:2014xxa}
that also occurs in the joint resummation of $q_T$ and $0$-jettiness~\cite{Procura:2014cba, Lustermans:2019plv}.

We can now change variables in \eq{factorization_collinear_endpoint_qm_qp_qT}
to $x_{a,b}$, using \eq{def_txab} and expanding in $\lambda$, which yields
\begin{align} \label{eq:factorization_collinear_endpoint_Q_Y_qT}
\frac{\df \sigma}{\df x_a \df x_b \, \df \vec{q}_T}
&= H_{ij}(Q^2) \int \! \df t  \, f_i^\thr\Bigl[x_a \Bigl(1+ \frac{q_T^2}{2Q^2} + \frac{t}{Q^2} \Bigr)\Bigr]
\nn\\ &\quad\times
B_j(t, \vec{q}_T, x_b)
\,.\end{align}
Crucially, when expanding $x_- = x_a[1 + q_T^2/(2Q^2) + \ord{\lambda^4}]$, we have
to keep the $q_T^2/(2Q^2)\sim \lambda^2$ term in the PDF argument because it
is of the same order as $t/Q^2 \sim \lambda^2$.
Integrating \eq{factorization_collinear_endpoint_Q_Y_qT} over $\vec{q}_T$,
we obtain
\begin{equation} \label{eq:factorization_collinear_endpoint_Q_Y}
\frac{\df \sigma}{\df x_a \df x_b}
= H_{ij}(Q^2) \!\int \! \df \tt \, f_i^\thr\Bigl[x_a\Bigl(1 + \frac{\tt}{Q^2} \Bigr)\Bigr]
\tB_j(\tt, x_b)
\,.\end{equation}
The factorization theorems in
\eqs{factorization_collinear_endpoint_Q_Y_qT}{factorization_collinear_endpoint_Q_Y}
hold at leading power in the generalized threshold limit $\lambda^2 \sim 1-x_a \ll 1$ for generic $x_b$.
They are our key new results.

In \eq{factorization_collinear_endpoint_Q_Y} we changed variables to
$\tt = t + q_T^2/2$, and defined the new modified beam function
\begin{align} \label{eq:def_modified_beam_function}
\tB_j(\tt, x) = \int \df^2\vec{k}_T\, B_j\Big(\tt - \frac{k_T^2}{2}, \vec{k}_T, x\Big)
\,.\end{align}
It has the same $\mu$ evolution as $B_j(t, x)$ but different constant terms.
It obeys a matching relation analogous to \eq{beam_function_matching}.
Using the known results for $B_j(t, \vec{k}_T, x)$~\cite{Jain:2011iu, Gaunt:2014xxa},
we have calculated its matching coefficients $\tcI_{jk}(\tt, z)$
to $\ord{\as^2}$ for $j=q$ and $\ord{\as}$ for $j = g$~\cite{supplement}.

The factorization structure in \eqs{factorization_collinear_endpoint_qm_qp}{factorization_collinear_endpoint_Q_Y} turns out to be analogous
to deep-inelastic scattering (DIS) at large Bjorken $x$~\cite{Sterman:1986aj, Catani:1989ne, Manohar:2003vb, Becher:2006mr, Fleming:2012kb, Chay:2013zya, Hoang:2015iva},
which factorizes as $H_{ij} f_i^{\rm thr}\otimes J_j$,
with the jet function $J_j$ describing collimated \emph{final-state} radiation.
For \eqs{factorization_collinear_endpoint_qm_qp}{factorization_collinear_endpoint_Q_Y}
to be consistent, the $\mu$ dependence of the functions
must cancel between them, and it does so in the same way as for DIS.
The beam function is known to have the same $\mu$ evolution as the
jet function~\cite{Stewart:2010qs}. The existence of the above factorization theorems
provides an independent confirmation of this.
The key differences to DIS are the additional dependence on $x_b$ and the
nontrivial $q_T$ dependence. The latter bears some resemblance to different
$1$-jettiness definitions in exclusive DIS~\cite{Kang:2013nha}.

Since \eq{factorization_collinear_endpoint_Q_Y} is valid for $x_a\to 1$ and arbitrary $x_b$,
it must contain the soft threshold factorization in \eq{factorization_soft} for $x_b\to 1$
as a special case. This implies~\cite{supplement}%
\begin{equation} \label{eq:consistency_collinear_endpoint_soft_threshold}
\tB_j(\omega k^-\!, x_b)
= \int \! \frac{\df k^+}{\omega}\, S(k^-, k^+)
f_j^\thr\Bigl[x_b\Bigl(1 + \frac{k^+}{\omega}\Bigr)\Bigr]
\end{equation}
to leading power in $1-x_b$, and identically for $B_j$.

We can now combine \eq{factorization_collinear_endpoint_Q_Y} with
the analogous result in the opposite limit $x_b\to 1$ for generic $x_a$ by
adding the two and subtracting their overlap, which is precisely given by the soft limit.
This yields our main result, the generalized threshold factorization theorem
\begin{align} \label{eq:factorization_generalized}
\frac{\df \sigma}{\df x_a \df x_b}
= H_{ij} \Bigl[ f_i^\thr\otimes\tB_j + \tB_i\otimes f_j^\thr - S \otimes f_i^\thr f_j^\thr \Bigr]
\,,\end{align}
with the convolutions as in \eqs{factorization_soft}{factorization_collinear_endpoint_Q_Y}.
Analogous results hold for $x_\pm$ and differential in $\vec{q}_T$.
Replacing $f_i^\thr[x(1 + 1-z)]$ by $f_i(x/z)/z$, which is justified at leading power in $1-z$,
and comparing to \eq{collinear_factorization},
we obtain the corresponding partonic factorization theorem
\begin{align} \label{eq:factorization_generalized_partonic}
\hsigma_{ij}(z_a, z_b)
&= H_{k\ell} \bigl[ \delta_{ki}\,\hcI_{\ell j}(z_a,z_b) + \hcI_{ki}(z_b, z_a) \, \delta_{\ell j}
\nn \\ &\quad\quad\quad
- \delta_{ki} \, \delta_{\ell j} \, \hS(z_a, z_b) \bigr]
\,,\end{align}
where we changed variables to $z_{a,b}$ and defined
\begin{align}
\hcI_{ij}(z_a, z_b) &\equiv Q^2\, \tcI_{ij}\bigl[Q^2(1-z_a), z_b \bigr]
\,, \nn \\
\hS(z_a, z_b) &\equiv Q^2\, S\bigl[ Qe^Y(1 - z_a), Qe^{-Y}(1-z_b) \bigr]
\,.\end{align}

\begin{figure}
\centering
\includegraphics[width=0.6\columnwidth]{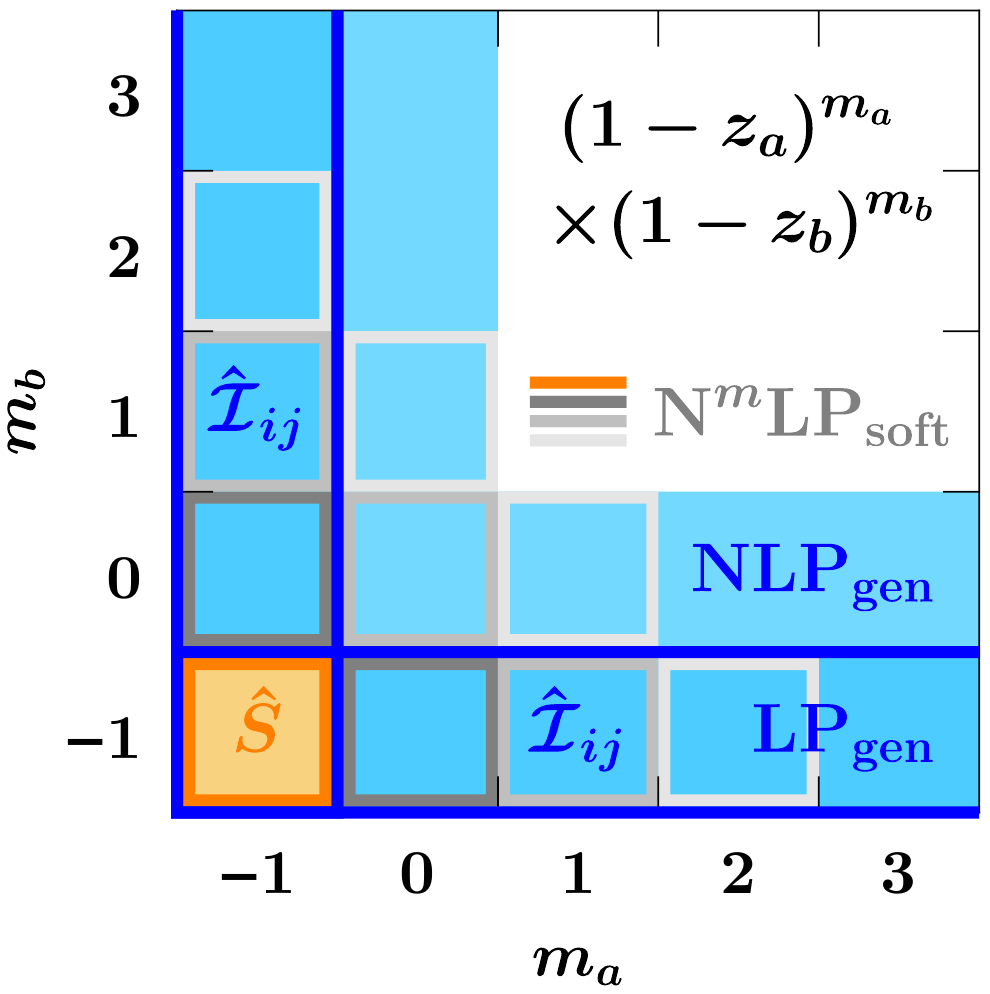}%
\caption{Terms in the partonic cross section $\hsigma(z_a, z_b)$
captured by the soft and generalized threshold expansions.}
\label{fig:power_expansion_partonic_xsec}
\end{figure}

As illustrated in \fig{power_expansion_partonic_xsec},
$\hcI_{ij}(z_a, z_b)$ captures all terms in $\hsigma(z_a, z_b)$ that are singular as $z_a \to 1$
with their exact dependence on $z_b$, and vice versa, with the overlap
given by $\hS(z_a, z_b)$. This includes flavor-nondiagonal contributions, e.g.,
$ij = qg$ with $k\ell = q\bq$ or $gg$ in \eq{factorization_generalized_partonic}.

\section{Validation}

\begin{figure}
\centering
\includegraphics[width=0.95\columnwidth]{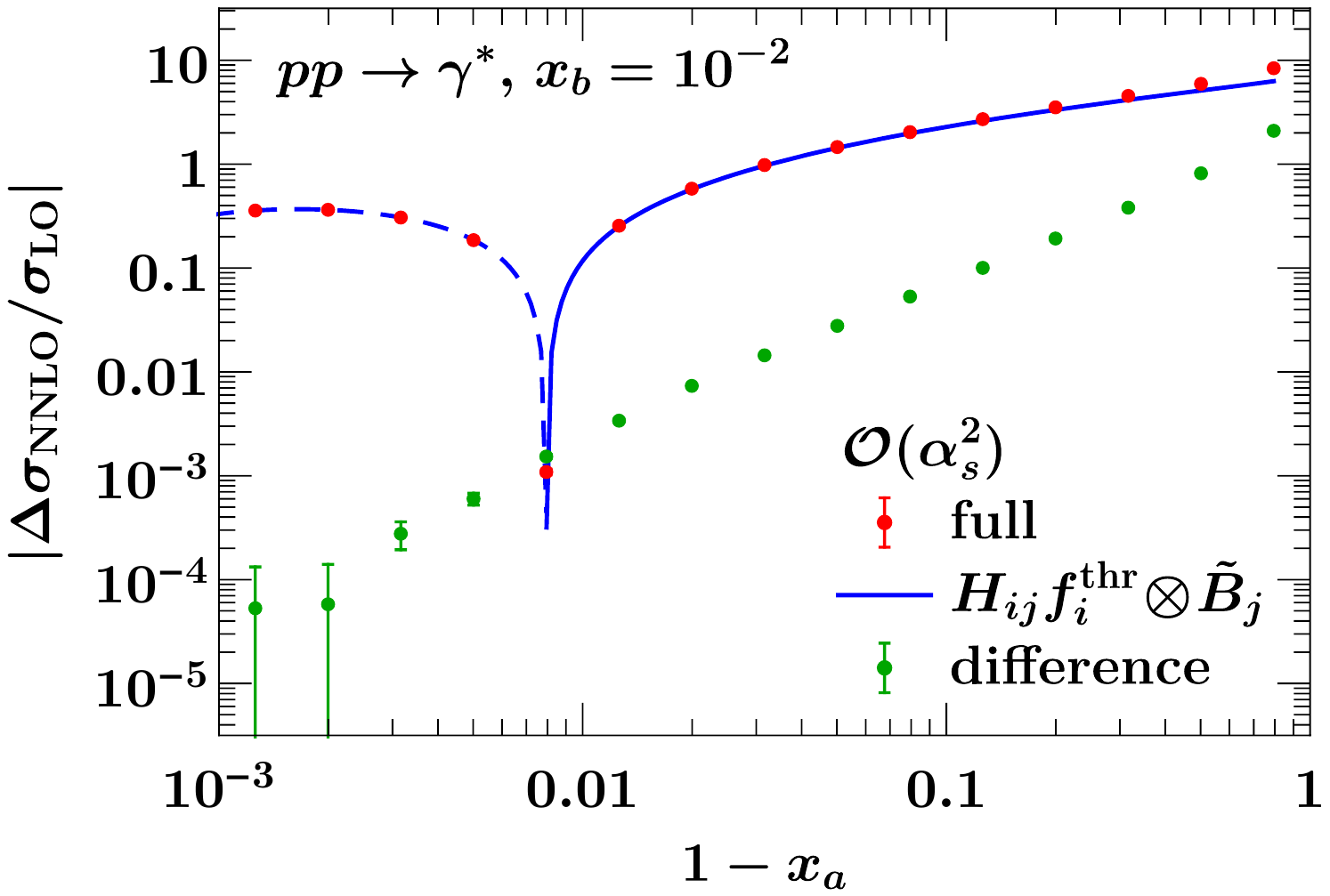}%
\caption{Validation of the $\ord{\as^2}$ contribution to $\df\sigma/\df x_a\df x_b$
predicted for $x_a \to 1$ by \eq{factorization_collinear_endpoint_Q_Y} (blue)
against the full result from \texttt{Vrap} (red).
Their difference (green) vanishes like a power as $1 - x_a \to 0$, as it must.
The error bars indicate the integration uncertainties.
}
\label{fig:sing_nons_gamma_nnlo_all_xb_1em2}
\end{figure}

A nontrivial validation is to check \eq{factorization_generalized}
against the full fixed-order result. First,
\eq{consistency_collinear_endpoint_soft_threshold} implies
\begin{align} \label{eq:consistency_collinear_endpoint_soft_threshold_partonic}
\hcI_{ij}(z_a, z_b)
&= \delta_{ij}\, \hS(z_a, z_b) \, \bigl[ 1 + \ord{1 - z_b} \bigr]
\,,\end{align}
which we verified analytically to $\ord{\as^2}$.
It then suffices to check that \eq{factorization_collinear_endpoint_Q_Y}
reproduces the fixed-order result for $x_a \to 1$ (or $z_a \to 1$)
for generic $x_b$ (or $z_b$).

At NLO, we can check analytically for Drell-Yan and $gg\to H$
against the results of \refscite{Anastasiou:2002qz, Anastasiou:2003yy},
which are given as distributions in $z = z_az_b$ and a variable $y(z_a, z_b)$.
Their translation to $(z_a, z_b)$ is nontrivial~\cite{supplement}.
We find complete agreement for all partonic channels.
We also combined all singular terms from \eq{factorization_generalized_partonic}
with the regular terms from \refscite{Anastasiou:2002qz, Anastasiou:2003yy} to construct
the full $\hsigma_{ij}(z_a, z_b)$ at NLO. The result is given in~\cite{supplement}
and agrees with \refscite{Kubar:1980zv, Mathews:2004xp, Ravindran:2006bu}.
\footnote{
   As discussed further in~\cite{supplement},
   several soft threshold factorizations differential in
   rapidity~\cite{Bolzoni:2006ky, Mukherjee:2006uu, Becher:2007ty,
   Bonvini:2010tp, Bonvini:2015ira}
   differ from \eq{factorization_soft} and do not reproduce the correct soft
   limit already at NLO.
}.

At NNLO, we numerically validate our own implementation of \eq{factorization_collinear_endpoint_Q_Y} in \texttt{SCETlib}~\cite{scetlib} against \texttt{Vrap}~\cite{Anastasiou:2003ds}.
We use flat PDFs, $f_i^\thr(x) = f_i(x) = \theta(1-x)$,
which amounts to taking cumulant integrals of the partonic cross section
and provides the strongest possible numerical check.
In \fig{sing_nons_gamma_nnlo_all_xb_1em2}, we compare the $\ord{\as^2}$ contribution
for Drell-Yan as a function of $1-x_a$ at fixed $x_b = 10^{-2}$.
We find perfect agreement.
The breakdown into partonic channels is given in~\cite{supplement}.
We also find similar agreement for other $x_b$ and for $pp \to Z/\gamma^\ast$
on the resonance.

\section{Illustrative applications}

\begin{figure}
\centering
\includegraphics[width=0.95\columnwidth]{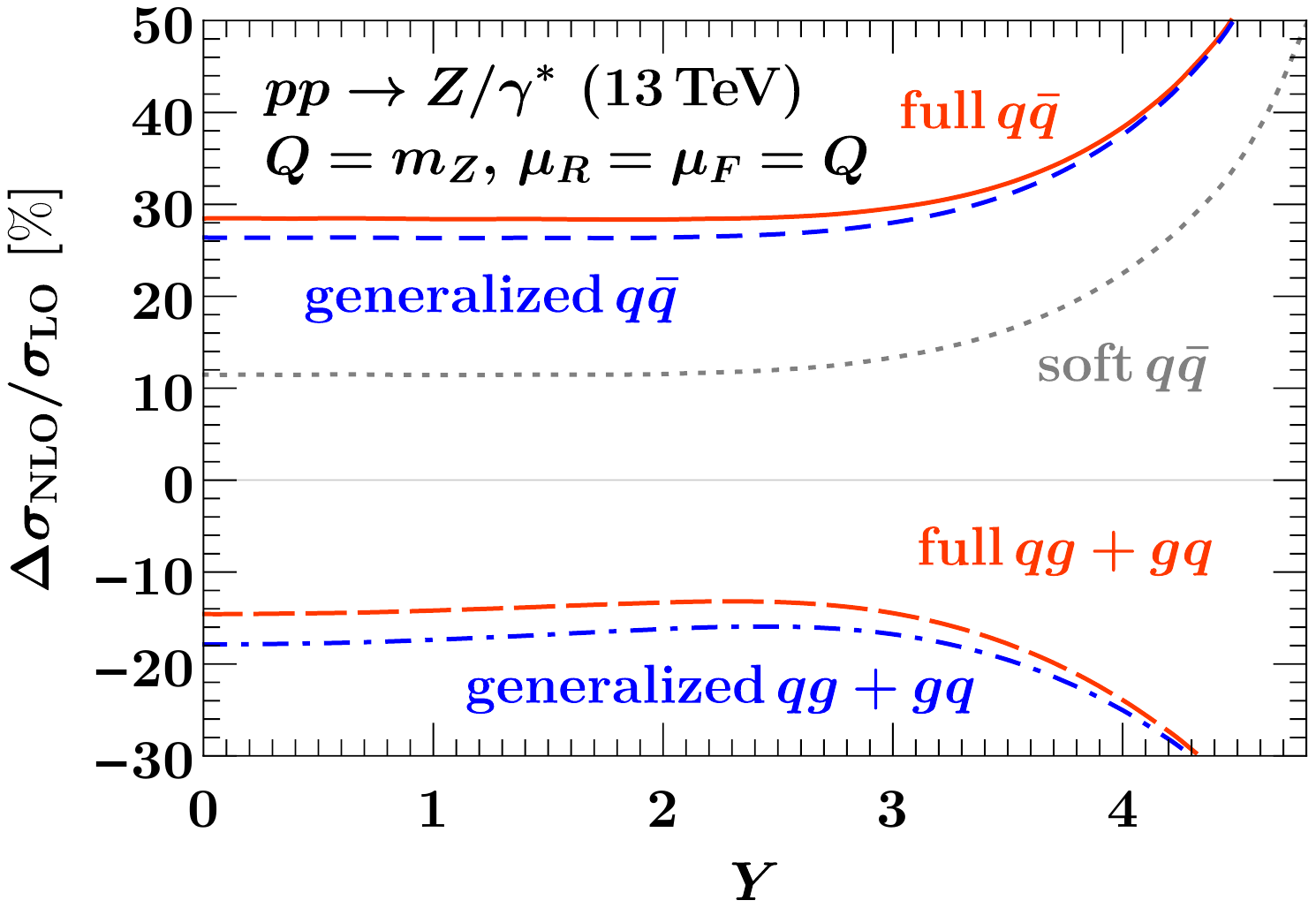}%
\\[1ex]
\includegraphics[width=0.95\columnwidth]{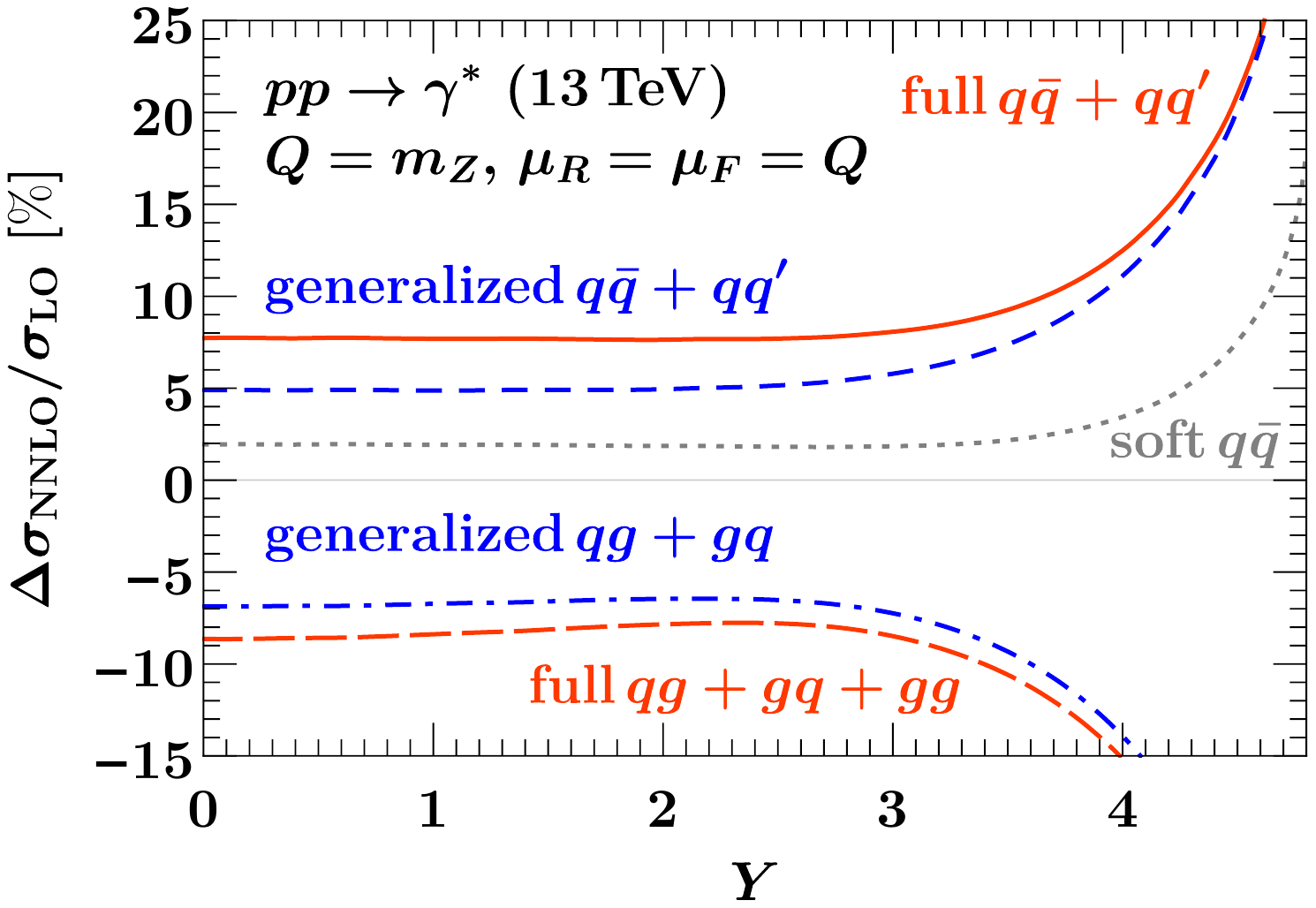}%
\caption{The $\ord{\as}$ (top) and $\ord{\as^2}$ (bottom)
contributions to $\sigma \equiv \df \sigma/\df Q \df Y$
normalized to the LO result.
Shown are the full result (red), the generalized threshold approximation (blue),
and the soft threshold approximation (gray).}
\label{fig:generalized_threshold_approximation_drell_yan}
\end{figure}

The immediate question that arises is how well the generalized threshold limit
approximates the full fixed-order result for physical PDFs, particularly in
comparison to the soft limit.
We use the \texttt{MMHT2014nnlo68cl}~\cite{Harland-Lang:2014zoa} PDFs,
\texttt{Vrap}~\cite{Anastasiou:2003ds} to obtain the full NNLO result%
\footnote{
   The public \texttt{Vrap 0.9} assumes
   $f_q(x) = f_{\bar q}(x)$ for $q = s,c,b$.
   We modified it to allow for different
   sea quark and antiquark PDFs.
}
and \texttt{SCETlib}~\cite{scetlib} to implement \eq{factorization_generalized_partonic}.
In \fig{generalized_threshold_approximation_drell_yan} we compare
the $\ord{\as}$ and $\ord{\as^2}$ contributions
to the Drell-Yan rapidity spectrum at $Q = m_Z$,
separated into quark channels ($q\bar{q} + qq'$)
and channels involving a gluon ($qg + gq + gg$).
Analogous results at $\mu = Q/2$ and for $gg\to H$ are provided in~\cite{supplement}.
The generalized threshold limit approximates the full result
well for all channels and all $Y$. As expected, it works particularly well toward large $Y$.
It works significantly better than the soft limit,
which only provides a poor approximation for the $q\bar q$ channel and none for
the others.
Currently, all ingredients are available to perform the resummation in the
generalized threshold limit to N$^3$LL, which we leave to future work.
We expect it to be much more powerful for improving the precision of
perturbative predictions than the soft threshold resummation.

Next, we may ask how well the power expansion around the generalized threshold limit
works and how it relates to the soft expansion.
Consider the double expansion in $1-z_a$ and $1 - z_b$, illustrated in \fig{power_expansion_partonic_xsec},
\begin{equation} \label{eq:power_expansion_general}
\hsigma_{ij}(z_a, z_b) = \sum_{m_a,m_b} \hsigma_{ij}^{(m_a,m_b)}(z_a, z_b)
\,,\end{equation}
where $\hsigma_{ij}^{(m_a,m_b)}(z_a, z_b) \sim (1-z_a)^{m_a}(1-z_b)^{m_b}$.
Expanding around the soft $z = z_a z_b \to 1$ limit corresponds to counting powers of
$(1-z)^{m_a + m_b + 1}$. The leading-power result in \eq{factorization_soft_partonic}
gives the $m_a = m_b = -1$ term. At the $m$th order, N$^m$LP$_\soft$, we keep all terms
with $m_a + m_b + 2 \leq m$.
At leading power in the generalized expansion, \eq{factorization_generalized_partonic}
includes all terms with $\min \{ m_a, m_b \} = -1$.
Similarly, at the $m$th order, N$^m$LP$_\gen$, we keep all terms with $\min \{ m_a, m_b \} = m - 1$,
so the \emph{missing} corrections at N$^m$LP$_\gen$ are $\ordsq{(1-z_a)^m(1-z_b)^m}$.

\begin{figure}
\centering
\includegraphics[width=0.95\columnwidth]{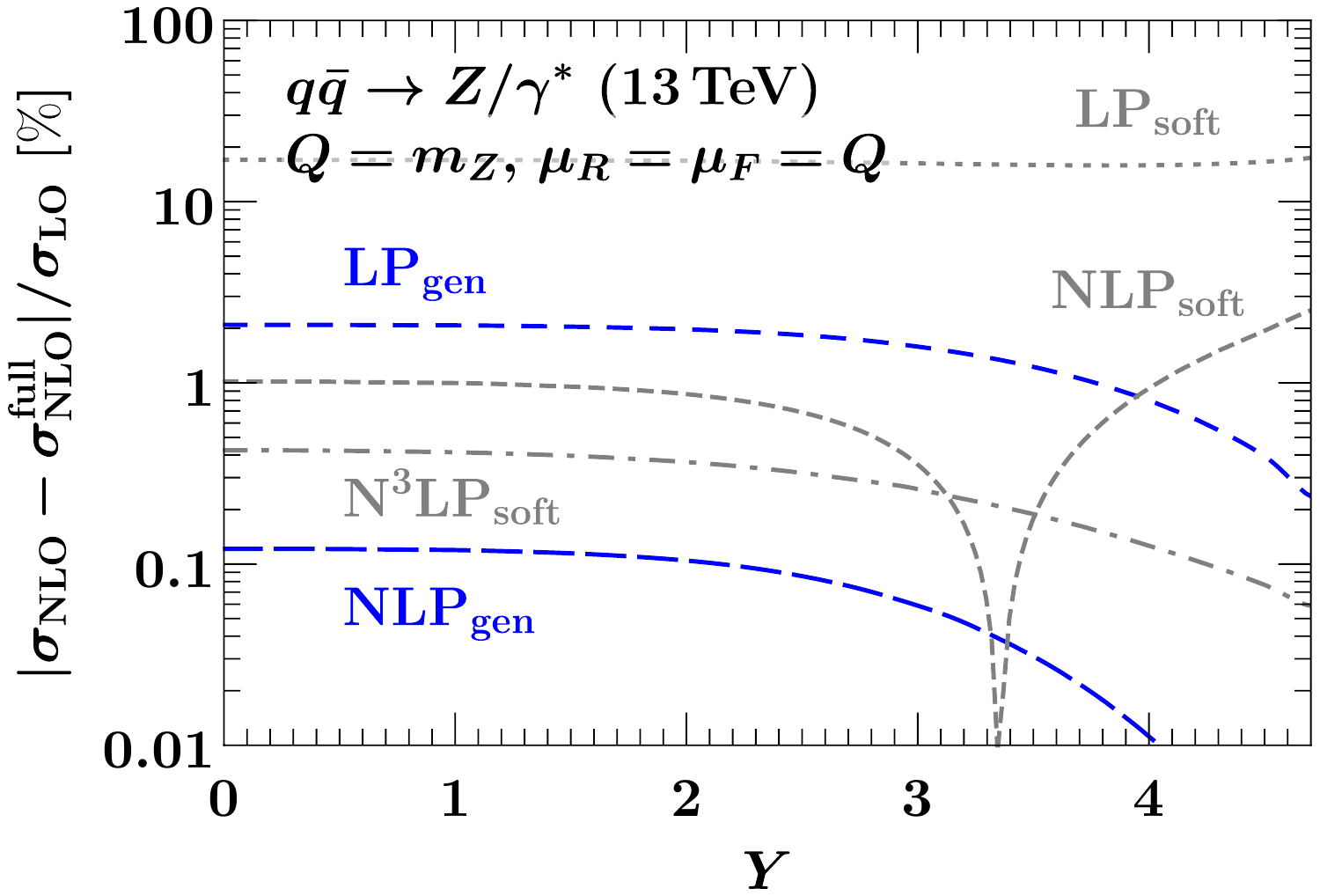}%
\\[1ex]
\includegraphics[width=0.95\columnwidth]{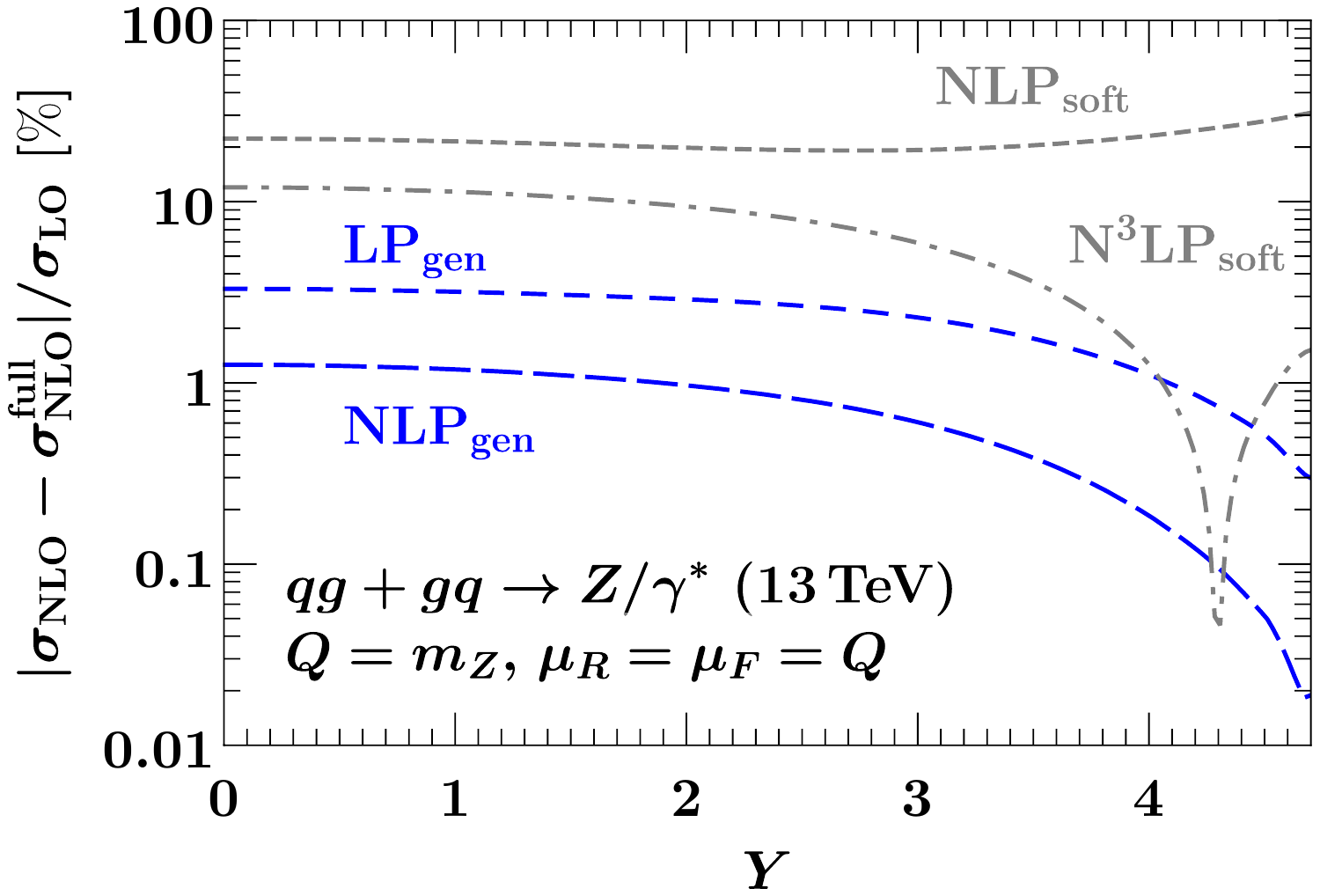}%
\caption{Convergence of the generalized (blue) and soft (gray) threshold
expansions. Shown are the deviations from the full NLO result
for the $q\bar{q}$ channel (top) and the $qg+gq$ channel (bottom)
normalized to the LO result.
\label{fig:generalized_threshold_approximation_drell_yan_convergence}}
\end{figure}

In \fig{generalized_threshold_approximation_drell_yan_convergence},
we show the deviation from the exact result at various orders in both expansions
for Drell-Yan at NLO, where we have full analytic control.
Analogous results for $gg\to H$ are provided in~\cite{supplement}.
The generalized expansion performs significantly better
than the soft one for both partonic channels.
We expect this to hold in general, since expanding a two-dimensional
function along a one-dimensional boundary is superior to expanding it in
a single point on that boundary.

In fact, as seen in \fig{power_expansion_partonic_xsec}, each order in the generalized
expansion fully contains two orders in the soft expansion, and in particular,
the LP$_\gen$ result \eq{factorization_generalized_partonic} contains
the entire NLP$_\soft$ contribution. This does not mean it can be used to perform the
NLP$_\soft$ resummation because the $\mu$ evolution
of $B_i(t, x)$ does not predict its $x$ dependence. It does, however,
show that $H_{ij}$ factorizes for all partonic channels
and reduces the problem to deriving the NLP$_\soft$
factorization for \eq{consistency_collinear_endpoint_soft_threshold}.

At N$^3$LO, \eq{factorization_generalized_partonic} predicts a highly
nontrivial set of terms for any color-singlet process, since all
terms $\sim \cL_n(1-z_a)$ in $\hcI_{ij}(z_a, z_b)$ are known from
its $\mu$ evolution, where $\cL_n(y) \equiv [\ln^n(y)/y]_+$.
To illustrate this, the coefficient of $\as^3/(4\pi)^3$ in $\hsigma_{ij}(\mu=Q)$
with $n\geq 3$ is
\begin{align} \label{eq:partonic_xsec_three_loops}
\hsigma_{ij}^{(3)} &=
\cL_5(1 - z_a) H^{(0)}_{ij} \delta(1-z_b) \, \frac{(\Gamma_0^i)^3}{8}
\\ & \quad
+ \cL_4(1 - z_a) H^{(0)}_{ik}
\biggl[ \delta_{kj} \delta(1-z_b) \Bigl( - \frac{2\beta_0 }{3} - \frac{\gamma_{B\,0}^i}{2} \Bigr)
\nn \\ &\quad
+ P^{(0)}_{kj}(z_b)
\biggr] \frac{5}{8}(\Gamma_0^i)^2
+ \cL_3(1-z_a) \biggl\{
   H^{(1)}_{ij} \delta(1-z_b) \frac{\Gamma_0^i}{2}
\nn \\ & \quad
   + H^{(0)}_{ik} \biggl[ \delta_{kj} \delta(1-z_b) \Bigl(\Gamma_1^i
   -\frac{\pi^2}{6} (\Gamma_0^i)^2 + \frac{\beta_0^2}{3}
\nn \\ &\quad
   + \frac{(\gamma_{B\,0}^i)^2}{4} + \frac{5}{6} \beta_0 \gamma_{B\,0}^i\Bigr)
- P^{(0)}_{kj}(z_b) \Bigl(\frac{5\beta_0}{3} + \gamma_{B\,0}^i \Bigr)
\nn \\ &\quad
+ \bigl( P^{(0)}_{k\ell} \otimes P^{(0)}_{\ell j} \bigr)(z_b)
+ \tI^{(1)}_{kj}(z_b) \, \frac{\Gamma_0^i}{2}
\biggl]
\biggr\} \Gamma_0^i + \dotsb
\nn\,.\end{align}
The required ingredients are given in~\cite{supplement}.
The extension down to $\cL_0(1-z_a)$ and to the full $z_a$ dependence for
$z_b\to 1$ is straightforward~\cite{Billis:2019xxx}. The $\delta(1-z_a)$ coefficient requires the still
unknown $\ord{\as^3}$ finite terms of the beam function.
For Drell-Yan, \eq{partonic_xsec_three_loops} significantly extends the current
knowledge at $\ord{\as^3}$~\cite{Ahmed:2014uya}, providing the full
$z_b$ dependence for all partonic channels. For $gg\to H$,
the extension to $\cL_0(1-z_{a})$ would also provide
additional information beyond what is currently known~\cite{Dulat:2018bfe}.

\section{Summary}

We introduced a generalized threshold factorization, which is much more powerful
than the often used soft threshold, thus opening the door to numerous
applications to improve theoretical predictions for collider processes.
It describes all kinematic limits in $(x_a, x_b)$ or $(Q, Y)$,
including in particular $\abs{Y} \to \Ymax$ at generic $Q$, which is directly accessible
at the LHC. It enables the corresponding threshold resummation where only one PDF is probed at
large $x$, which is not captured by the soft limit.
At the partonic level, it captures all singularities of $\hsigma_{ij}(z_a, z_b)$,
including nondiagonal partonic channels.
It can be used to predict a rich set of terms at higher fixed
order or to resum them to all orders.
It is the weakest known limit in which the process-dependent virtual corrections
($H_{ij}$) factorize.

While we only considered color-singlet
processes here, the same methods can be used to generalize the
soft threshold factorization in other situations, such as for processes with
heavy particles, jets, or identified hadrons in the final state.

\begin{acknowledgments}
\paragraph{Acknowledgments}
We thank M.~Beneke, G.~Billis, A.~Broggio, G.~Das, M.~Diehl, L.~Dixon, M.~Ebert,
S.~Forte, A.~Kulesza, S.~Marzani, B.~Mistlberger, F.~Ringer, D.~Scott, D.~Soper,
and W.~Waalewijn for discussions.
We thank M. Stahlhofen for providing the results of \refcite{Gaunt:2014xxa},
and M.\ Prakash and V.\ Ravindran for help in comparing with
\refscite{Mathews:2004xp, Ravindran:2006bu}.
The authors thank each other's institutions for hospitality.
This work was supported in part by the ERC grant ERC-STG-2015-677323 and the
D-ITP consortium, a program of the Netherlands Organization for Scientific Research
(NWO) that is funded by the Dutch Ministry of Education, Culture and Science (OCW).
\end{acknowledgments}

\bibliography{references}


\onecolumngrid
\clearpage

\setcounter{equation}{1000}
\setcounter{figure}{1000}
\setcounter{table}{1000}

\renewcommand{\theequation}{S\the\numexpr\value{equation}-1000\relax}
\renewcommand{\thefigure}{S\the\numexpr\value{figure}-1000\relax}
\renewcommand{\thetable}{S\the\numexpr\value{table}-1000\relax}

\section*{Supplemental material}

\subsection{Factorization theorem}

\begin{table}[t]
\renewcommand{\arraystretch}{1.5}
\tabcolsep 5pt
\centering
\begin{tabular}{c||c|c}
\hline\hline
Mode & Lab frame & Leptonic ($\hat Y = 0$) frame
\\[-1ex]
& $(+,-,\perp)$ & $(+,-,\perp)$
\\ \hline
$p_\bn$ & $(q^+, \lambda^2 q^-, \lambda \sqrt{q^- q^+})$ & $Q\, (1, \lambda^2, \lambda)$
\\[0.5ex]
$P_\bn$ & $\displaystyle\Bigl(q^+, \frac{\LQCD^2}{q^+}, \LQCD\Bigr)$ & $Q\, (1, \lqcd^2, \lqcd)$
\\[1.5ex]\hline
&&\\[-3ex]
$P_n$ & $\displaystyle\Bigl(\frac{\LQCD^2}{q^-}, q^-, \LQCD\Bigr)$ & $Q\,(\lqcd^2, 1, \lqcd)$
\\[1.5ex]
$P_s$ & $\displaystyle\Bigl(\frac{1}{\lambda^2}\,\frac{\LQCD^2}{q^-}, \lambda^2 q^-, \LQCD \Bigr)$
& $\displaystyle Q\,\Bigl(\frac{\lqcd^2}{\lambda^2}, \lambda^2, \lqcd\Bigr)$
\\[1.5ex]\hline
&&\\[-3ex]
$P_{us}$ & $\displaystyle\Bigl(\frac{\LQCD^2}{q^-}, \frac{\LQCD^2}{q^+}, \frac{\LQCD^2}{\sqrt{q^+ q^-}}\Bigr)$
& $Q\, (\lqcd^2, \lqcd^2, \lqcd^2)$
\\[1.5ex]
$P_G$ & $\displaystyle\Bigl(\frac{\LQCD^2}{q^-}, \frac{\LQCD^2}{q^+}, \LQCD \Bigr)$
& $Q\, (\lqcd^2, \lqcd^2, \lqcd)$
\\[1ex]\hline\hline
\end{tabular}
\caption{Relevant EFT modes in the limit $\lqcd \sim \lambda^2 \sim 1 - x^- \ll 1$
in the lab (hadronic center-of-mass) frame and the leptonic frame where $\hat Y = 0$.
In the right column we used that in the leptonic frame, $q^\pm \to \hat q^\pm = \sqrt{q^+ q^-} \sim Q$.}
\label{tab:modes}
\end{table}

Here we give some more details on the derivation of the factorization theorem
in \eq{factorization_collinear_endpoint_qm_qp_qT}, which underlies all other factorization
theorems, and which we repeat here for easy reference,
\begin{align} \label{eq:factorization_collinear_endpoint_qm_qp_qT_again}
\frac{\df \sigma}{\df x_- \df x_+ \, \df \vec{q}_T}
&= H_{ij}(q^+ q^-) \int \! \df t  \, f_i^\thr \Bigl[x_-\Bigl(1 + \frac{t}{q^+ q^-} \Bigr)\Bigr]
B_j(t, \vec{q}_T, x_+)
\,.\end{align}
As in the main text, we always implicitly sum over repeated flavor indices $i,j,k$.
Equation~\eqref{eq:factorization_collinear_endpoint_qm_qp_qT_again}
is valid up to power corrections in $\lambda^2$ in the generalized threshold limit
\begin{equation} \label{eq:gen_threshold_limit}
\lambda^2 \sim 1 - x_- \ll 1
\qquad\text{for generic } x_+
\,,\qquad
\lqcd \equiv \LQCD/Q \sim \lambda^2 \ll \lambda
\,.\end{equation}
We require $\lqcd \ll \lambda$ for reasons that will be apparent soon.
Without loss of generality we can then consider $\lqcd \sim \lambda^2$. This relation
is to be interpreted as follows:
First, in our context, $\lqcd$ denotes the scale of the PDFs, which is generically allowed to be as large as
$\lambda^2$ and does not necessarily have to be nonperturbative. If it happens to be
a perturbative scale, then the physics below $\lqcd$ is simply described by the PDF evolution.
Conversely, it also means that $\lambda^2$ is in principle allowed to be
as small as $\lqcd$ including being nonperturbative, i.e., it is only relevant that
$\lambda\gg\lqcd$ is perturbative.

The key step in deriving \eq{factorization_collinear_endpoint_qm_qp_qT_again}
is to identify the relevant degrees of freedom (modes) in the effective field theory (EFT)
that describe the physical situation. They are defined by the relative
scaling of their light-cone momentum components and are summarized in \tab{modes}.
We note that rather than matching QCD directly onto these modes,
one may also perform a multi-stage matching, as was done for endpoint DIS in \refcite{Hoang:2015iva},
which yields the same end result.

The $p_{\bn}$ modes describe the hadronic final state of the collision.
Their scaling is determined by the fact that in the limit of \eq{gen_threshold_limit}, there is only
$p_\bn^- \sim \lambda^2 \Ecm \sim \lambda^2 q^-$ minus momentum available.
On the other hand, their plus momentum is unconstrained, which means it has
generic scaling set by the hard interaction, $p_\bn^+ \sim \xi_b \Ecm \sim x_b \Ecm \sim q^+$.
Since $p_\bn^2 \sim \lambda^2 q^+ q^- \sim \lambda^2 Q^2 \gg \LQCD^2$, the $p_\bn$ modes
are perturbative. Therefore, they describe the perturbative QCD final state produced in the
partonic collision in addition to $L$ (but excluding the beam remnant).
The $P_n$ and $P_\bn$ modes describe the incoming protons, or more precisely,
the partons in the proton with the typical momentum fractions required to produce the
hard final state. This means their scaling is determined by $P_n^- \sim q^-$ and
$P_\bn^+ \sim \xi_b \Ecm \sim x_b \Ecm \sim q^+$ and $P_n^2 \sim P_\bn^2 \sim \LQCD^2$.

The soft $P_s$ modes describe the interactions between the $p_\bn$ and $P_n$ modes.
Their scaling is thus determined by $P_s^- \sim p_\bn^- \sim \lambda^2 Q$
and $P_{s\perp}\sim P_{n\perp} \sim \LQCD$ or equivalently $P_s^2 \sim P_n^2 \sim \LQCD^2$.
Hence, they keep the $p_\bn$ modes on shell and have a SCET$_{\rm I}$-like relation to
them. Their interactions with the $p_\bn$ modes in the leading-power SCET Lagrangian are
decoupled and moved into soft Wilson lines in the SCET current via the BPS field redefinition.
At the same time, the $P_s$ modes have a SCET$_{\rm II}$-like relation to the $P_n$ modes, i.e.,
they have the same virtuality but are parametrically separated in rapidity. Hence,
their interactions with the $P_n$ modes, which take the $P_n$ modes off shell, are
described by soft Wilson lines in the SCET current that are directly produced
during the matching onto SCET. The distinction of the $P_s$ modes relies
on $\lqcd \ll \lambda$, while for $\lqcd\sim\lambda$, they would
become degenerate with the $p_\bn$ and $P_\bn$ modes.

The power counting and the relations between the modes are easiest in the leptonic frame,
which is the frame where the color-singlet final state has total rapidity $\hat Y = 0$.
Boosting from the lab frame to the leptonic frame by $Y$, we have
$\hat q^\pm = q^\pm e^{\pm Y} = \sqrt{q^+ q^-} \sim Q$. In the leptonic frame, the $p_\bn$
modes are genuinely $\bn$-collinear with $p_\bn^- \sim \lambda^2 p_\bn^+$, and
the soft modes are homogeneous, $P_s \sim \lambda^2 Q \sim \lqcd Q$.
By contrast, in the lab frame we must separately keep track of $q^+$ and $q^-$, i.e.,
we cannot count them as $q^+ \sim q^-$, because we want to take the limit of large $q^-$
for generic $q^+$. As a result, the $p_\bn$ modes do not necessarily appear to be $\bn$-collinear
in the lab frame because $q^+ \sim \lambda^2 q^-$ is allowed.
However, the key requirement for their factorization is that they are collinear \emph{relative}
to the soft modes, which in the lab frame are boosted in the $n$-collinear direction
and become $n$-collinear-soft (csoft) modes~\cite{Bauer:2011uc, Procura:2014cba}.

Finally, the ultrasoft (usoft) $P_{us}$ and Glauber $P_{G}$ modes describe the interactions
between the $P_n$ and $P_\bn$ modes that are possible without pushing either of them off shell,
which requires $P_{us}^- \sim P_\bn^-$ and $P_{us}^+ \sim P_n^+$. The $\perp$ component
of the usoft modes is fixed by requiring them to be on-shell modes, $P_{us\perp}^2 \sim P_{us}^+ P_{us}^-$.
The corresponding Glauber modes are allowed to be off shell, so their $\perp$ component
can be as large as $P_{G\perp}\sim P_{n\perp}\sim P_{\bn\perp}\sim \LQCD$.
The effects of the usoft and Glauber modes cancel, so we do not need to consider them
further. This directly follows from the collinear
factorization theorem~\cite{Bodwin:1984hc, Collins:1985ue, Collins:1988ig} because
in the limit we consider, the measurement is still fully inclusive over
perpendicular momenta at the scale $\lqcd Q$. This is another reason why we require
$\lqcd \ll \lambda$.
Note also that there is only a single collinear sector at the perturbative
$\lambda Q$ scale, so there are no perturbative Glauber modes with scaling
$Q(\lambda^2, \lambda^2, \lambda)$ that could spoil factorization.

Since there are no interactions between the modes in the leading-power SCET Lagrangian,
the cross section factorizes into separate forward matrix elements in each sector.
The detailed derivation closely follows \refcite{Stewart:2009yx}, with the
matrix element of the combined $p_\bn$ and $P_\bn$ modes giving
$B_j(t, x)$ and the combined matrix element of the $P_n$ and $P_s$ modes giving
$f_i^\thr$. The factorization of the threshold PDF into separate collinear and csoft
nonperturbative matrix elements is discussed in \refscite{Fleming:2012kb, Hoang:2015iva}.
It is not needed for our purposes.
The arguments of the functions and their convolution structure in
\eq{factorization_collinear_endpoint_qm_qp_qT_again} follow from overall
momentum conservation among all sectors, which at leading power in $\lambda$
must hold separately for the large (label) and small (residual) momenta carried
by the matrix elements. Denoting them as $\omega_{n,\bn}\sim Q$,
$k_{\bn\perp}\sim \lambda Q$, and $k_{s,\bn}\sim \lambda^2 Q$, it takes the form
\begin{equation} \label{eq:mom_cons}
\delta(\omega_\bn - q^+)\, \delta(\omega_n - q^-)
\,\delta(\vec k_{\bn \perp} + \vec q_T)
\,\delta(k_s^- - k_\bn^-)
\,.\end{equation}
The first three $\delta$ functions set the $x_\pm$ and $\vec{q}_T$ arguments of
$f_i^{\rm thr}$ and $B_j$.
The analogous $\delta(k_s^+ - k_\bn^+)$ disappeared by absorbing the $k_\bn^+$ dependence into $\omega_\bn$~\cite{Stewart:2009yx}.
By combining the $P_n$ and $P_s$ modes, the threshold PDF depends on
$(\omega_n + k_s^-)/P_a^- = x_-(1 + k_s^-/q^-)$.
Here, we do have to keep track of the momentum of the $P_s$ modes $k_s^-\sim\lambda^2Q$,
as it is much larger than the typical residual minus momentum $k_n^- \sim \lqcd^2Q$
of the $P_n$ modes that is absorbed into $\omega_n$.
The $\delta(k_s^- - k_\bn^-)$ then yields the convolution in $t = q^+ k_\bn^-$
in \eq{factorization_collinear_endpoint_qm_qp_qT_again}.
While \eq{factorization_collinear_endpoint_qm_qp_qT_again} is formally derived
in the leptonic frame, it has exactly the same form in the lab frame.
This is because the measured observables $x_\pm$ and $\vec{q}_T$ are boost invariant along the
beam axis, and reparametrization invariance (RPI)~\cite{Manohar:2002fd}
forces all functions to only depend on boost-invariant quantities.

To see that the generalized threshold limit contains the soft threshold limit,
note that the limits $x_- \to 1$ and $x_+ \to 1$ commute,
so taking one limit after the other is equivalent to taking $x_-, x_+ \to 1$ simultaneously.
To see this, consider the hierarchy $\lambda_-^2  \sim 1-x_- \ll \lambda_+^2 \sim 1-x_+ \ll 1$,
which we can interpret as taking $x_+\to 1$ after having already taken $x_- \to 1$.
In this limit, the $p_\bn$ modes factorize into perturbative $\bn$-csoft modes
$p_{\bn,cs} \sim Q(\lambda_+^2, \lambda_-^2, \lambda_+\lambda_-)$ and
$\bn$-csoft modes $P_{\bn,cs} \sim Q(\lambda_+^2, \lqcd^2/\lambda_+^2, \lqcd)$.
Including $\lambda_+$, the condition $\lqcd \ll \lambda$ becomes $\lqcd \ll\lambda_-\lambda_+$
and without loss of generality we can consider $\lqcd \sim \lambda_-^2\lambda_+^2$.
This also implies that the $P_s$ modes now get boosted in the $n$ direction
and become $n$-csoft modes, $P_s \equiv P_{n,cs}$.
The beam function matching onto PDFs now takes the form
\begin{equation} \label{eq:beam_soft_limit}
B_j(\omega_\bn k^-, \vec{k}_T, x_+)
= \int \! \frac{\df k^+}{\omega_\bn} \, \mathcal{S}(k^-, k^+, \vec{k}_T) \,
f_j^\thr\Bigl[x_+\Bigl(1 + \frac{k^+}{\omega_\bn}\Bigr)\Bigr] \,
\Bigl[ 1 + \ord{\lambda_+} \Bigr]
\,,\end{equation}
where the combined $P_{\bn,cs}$ and $P_\bn$ modes yield the threshold PDF, and
$\mathcal{S}(k^-, k^+, \vec{k}_T)$ is the matrix element of the $p_{\bn,cs}$ modes.
It has the same Wilson line structure
as the soft function appearing in \eq{factorization_soft},
and thus upon integration over $\vec{k}_T$ becomes equal to it to all orders
by reparametrization invariance.
We now have $k_T^2/Q^2 \sim \lambda_-^2 \lambda_+^2 \ll t/Q^2 \sim \lambda_-^2$,
so $B_j$ and $\tB_j$ become the same and integrating \eq{beam_soft_limit}
over $\vec{k}_T$ yields \eq{consistency_collinear_endpoint_soft_threshold} for
either of them.

\subsection{Plus distribution identities}

\begin{figure*}
\centering
\includegraphics[width=0.4\textwidth]{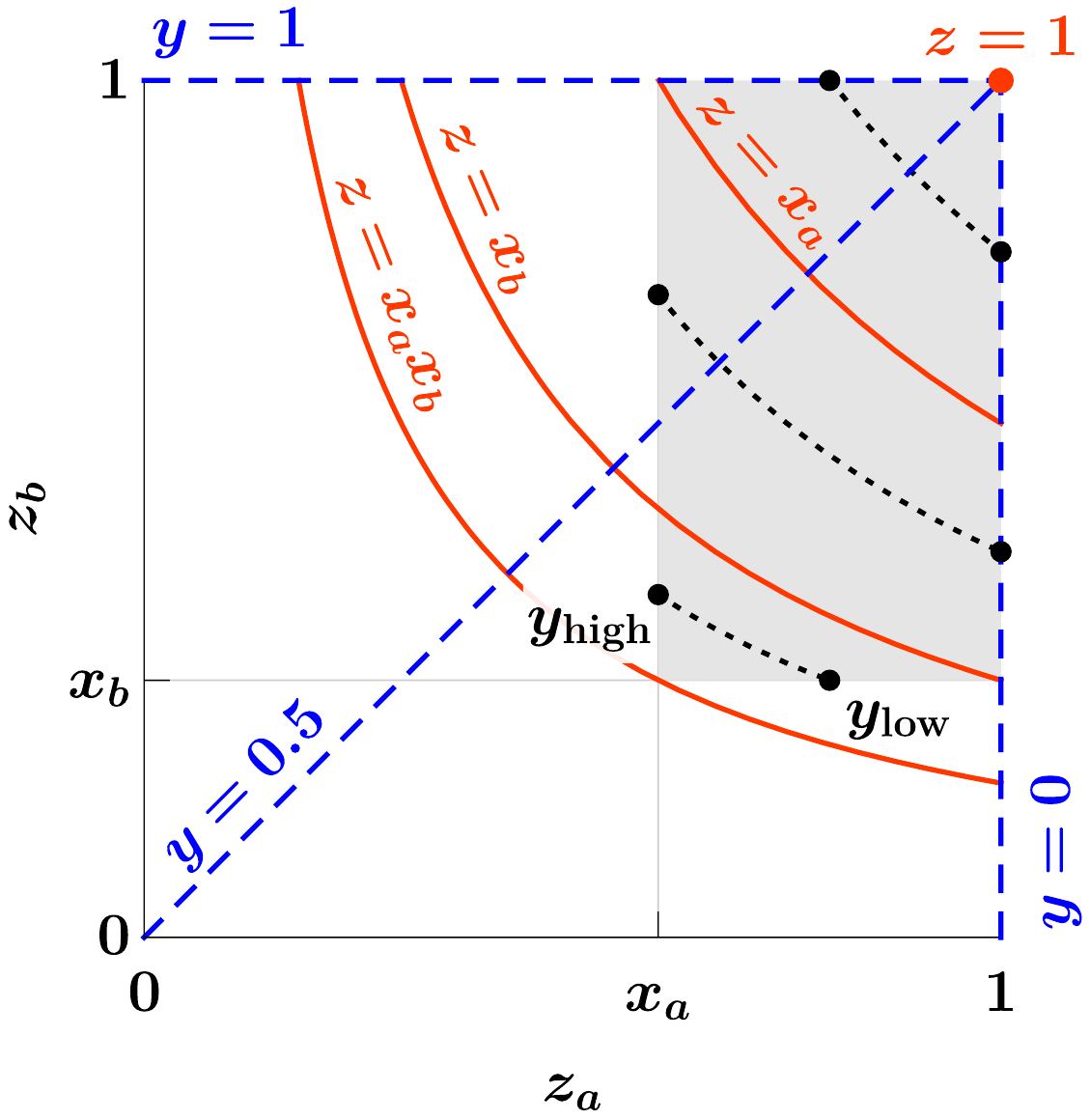}%
\caption{The $(z_a, z_b)$ plane as parametrized by $(z,y)$.
The gray area shows the integration region $x_a \leq z_a \leq 1$ and $x_b \leq z_b \leq 1$
(for the case of $x_a > x_b$) used to derive the distribution identities.
The dotted black lines indicate integration paths over $y$ at representative fixed values of $z$.
The solid red lines indicate the edge cases in $z$.
The dashed blue lines are lines of constant $y$.
\label{fig:integration_region}}
\end{figure*}

In \refscite{Anastasiou:2002qz, Anastasiou:2003yy, Anastasiou:2003ds}, the
partonic cross section for $(Q, Y)$ is given in terms of partonic variables $(z, y)$, defined as
\begin{alignat}{7} \label{eq:definition_z_y}
z &= z_az_b
\,, \qquad &
y
&= \frac{z_b(1 - z_a^2)}{(1-z_az_b)(z_a+z_b)}
\,, \qquad &
1-y
&= \frac{z_a(1 - z_b^2)}{(1-z_az_b)(z_a+z_b)}
= y|_{a \leftrightarrow b}
\,, \nn \\
z_a &= \sqrt{\frac{z[1-y(1-z)]}{z+y(1-z)}}
\,, \qquad &
z_b &= z_a|_{y \leftrightarrow 1-y}
\,, \qquad &
\frac{\df z \, \df y}{\df z_a \df z_b}
&= \frac{2[1-y(1-z)] [1 - (1-y)(1-z)]}{1-z^2}
\,,\end{alignat}
where $z_{a,b}$ are defined by \eq{collinear_factorization}, and the integration
limits $0 \leq z_{a,b} \leq 1$ correspond to $0 \leq z \leq 1$ and $0 \leq y \leq 1$.
In the following, we derive relations between plus distributions in $(z,y)$ and
$(z_a, z_b)$.

In general, the plus distributions are uniquely defined by their functional form in the bulk,
i.e.\ for $z_{a,b} < 1$ away from any singularity, and their integrals (against unit
test functions) over arbitrary integration regions that include the singularities.
First, the functional form in the bulk is easily obtained simply by plugging in
\eq{definition_z_y}.
Next, the correct boundary terms at $z_a = 1$ as a function of $z_b$,
at $z_b = 1$ as a function of $z_a$, and at $z_a = z_b = 1$ are
determined by comparing integrals over the integration region $x_a \leq z_a(z,y) \leq 1$
and $x_b \leq z_b(z,y) \leq 1$ for generic $x_a$ and $x_b$.
This integration region is indicated by the gray box in the $(z_a,z_b)$ plane in \fig{integration_region}.
It is sufficiently general to fix all boundary terms and precisely corresponds to the
relevant integration region for the physical cross section in $(Q, Y)$.
In terms of $(z,y)$, the integration region is given by
\begin{align} \label{eq:translation_integration}
\theta\bigl[z_a(z,y) \geq x_a\bigr] \, \theta\bigl[z_b(z,y) \geq x_b\bigr]
& = \theta\bigl[x_ax_b \leq z < \min\{x_a,x_b\}\bigr] \, \theta\bigl[y_\low(z) \leq y \leq y_\high(z)\bigr]
+ \theta \bigl[ \max\{x_a,x_b\} \leq z \bigr]
\nn \\ & \quad
+ \theta\bigl[x_b \leq z < x_a\bigr] \, \theta\bigl[y \leq y_\high(z)\bigr]
+ \theta\bigl(x_a \leq z < x_b) \, \theta \bigl[y_\low(z) \leq y \bigr]
\,,\end{align}
where the integration bounds in $y$, also illustrated in \fig{integration_region}, are given by
\begin{align}
y_\high(z) = \frac{z(1-x_a^2)}{(1-z)(x_a^2+z)}\,, \qquad y_\low(z) = \frac{x_b^2-z^2}{(1-z)(z+x_b^2)}
\,.\end{align}

In \tab{distribution_dictionary_z_y}, we collect the distribution identities
at leading power in $1-z_a$ and arbitrary $z_b$ that are required for validating
\eq{factorization_collinear_endpoint_Q_Y} at NLO, where we denote the plus distributions as
\begin{align}
\cL_n(x) = \Bigl[ \frac{\theta(x) \ln^n(x)}{x} \Bigr]_+ = \lim_{\epsilon \to 0} \frac{\df}{\df x} \Bigl[ \theta(x-\epsilon) \frac{\ln^{n+1} x}{n+1} \Bigr]
\quad\text{with}\quad
\cL_n(x > 0) = \frac{\ln^n(x)}{x}
\,, \qquad
\int^1 \! \df x \, \cL_n(x) = 0
\,.\end{align}
The relations are derived by integrating each structure in the left column in terms of
$(z, y)$ over the region in \eq{translation_integration}, expanding the result to
leading power in $1-x_a$, and comparing to the corresponding (straightforward) integral
over the same region in terms of $z_{a,b}$ of each structure in the right column.

\begin{table}[t]
\renewcommand{\arraystretch}{1.5}
\tabcolsep 5pt
\centering
\begin{tabular}{ll}
\hline\hline
\multicolumn{1}{l}{$\df z \, \df y \, \text{LHS}$}
&\multicolumn{1}{l}{$ = \df z_a \df z_b \, \text{RHS}$} \\
\hline
$f(z) \, \delta(y)$
&$f(z_b)\, \delta(1-z_a)$
\\
$r(z) \, \cL_0(y)$
& $\displaystyle r(z_b) \Bigl[\cL_0(1-z_a)
+ \delta(1-z_a)\, \ln \frac{2z_b}{(1+z_b)(1-z_b)} \Bigr] + \ord{1}$
\\[1ex] \hline
$f(z) \, \delta(1-y)$
& $f(z_a) \delta(1-z_b)$
\\
$r(z) \, \cL_0(1-y)$
& $\ord{1}$
\\ \hline
$r(z) \, \ordsq{y^0(1-y)^0}$
& $\ord{1}$
\\ \hline
$\cL_0(1-z) \, \bigl[\cL_0(y) + \cL_0(1-y) \bigr]$
& $\displaystyle- \cL_1(1-z_a)\, \delta(1-z_b)
+ \cL_0(1-z_a)\, \cL_0(1-z_b) - \delta(1-z_a)\, \cL_1(1-z_b)$
\\
& $\displaystyle + \frac{\pi^2}{6} \delta(1-z_a)\, \delta(1-z_b)  + \delta(1-z_a)\, \frac{1}{1-z_b} \ln \frac{2z_b}{1+z_b} + \ord{1}$
\\[1ex]\hline\hline
\end{tabular}%
\caption{Translation identities of two-dimensional plus distributions between the
$(z, y)$ and $(z_a, z_b)$ parametrizations.
Here, $f(x)$ is an arbitrary function of $x$, potentially distribution-valued for $x\to 1$,
while $r(x) = \ordsq{(1-x)^0}$ has at most an integrable singularity for $x\to 1$.
When indicated, the relations receive power corrections in $1-z_a$ starting at $\ord{1} \equiv \ordsq{(1-z_a)^0}$.
Overall factors of $\theta(1-z) \, \theta(y) \, \theta(1-y) = \theta(1-z_a) \, \theta(1-z_b)$ are understood on both sides.
}
\label{tab:distribution_dictionary_z_y}
\end{table}

The last entry in \tab{distribution_dictionary_z_y} has the most intricate structure.
Its exact integral without any expansion is given by
\begin{align}
&\int \! \df z \df y\, \theta\bigl[z_a(z,y) - x_a\bigr]\,\theta\bigl[z_b(z,y) - x_b\bigr] \, \cL_0(1-z) \Bigl[ \cL_0(y) + \cL_0(1-y) \Bigr]
\nn \\ & \qquad
= F_a(x_a) + F_b(x_b) - F_a(x_ax_b) - F_b(x_ax_b)
\,, \nn \\
\text{with}\qquad
F_{a,b}(z) &= - \ln(1-z)\ln(-z+\img 0) - \text{Li}_2\Big(\frac{1-z}{1+x_{a,b}}\Big)  - \text{Li}_2\Big(\frac{1-z}{1-x_{a,b}} - \img 0\Big)  - \text{Li}_2(z)
\,,\end{align}
where the imaginary parts from the branch cuts cancel between the different terms.
Matching this with the exact distribution in the bulk, we obtain the distributional identity
\begin{align} \label{eq:translation_L0L0_exact}
&\df z \, \df y \, \cL_0(1-z) \, \bigl[\cL_0(y) + \cL_0(1-y)\bigr]
\nn \\ & \qquad
= \df z_a \df z_b \biggl[
\frac{\pi^2}{6}\, \delta(1-z_a)\, \delta(1-z_b) - \cL_1(1-z_a)\, \delta(1-z_b)
+ \cL_0(1-z_a)\, \cL_0(1-z_b) - \delta(1-z_a)\, \cL_1(1-z_b)
\nn \\ &\qquad\qquad\qquad\quad
+ \delta(1-z_a)\, \frac{1}{1-z_b} \ln \frac{2z_b}{1+z_b} + \delta(1-z_b)\, \frac{1}{1-z_a} \ln \frac{2z_a}{1+z_a}
+ \frac{1}{(1 + z_a)(1 + z_b)}
\biggr]
\,.\end{align}
Expanding the right-hand side to leading power in $1-z_a$ yields the result
given in the last line of \tab{distribution_dictionary_z_y}.
We note that \eq{translation_L0L0_exact} is the two-dimensional analog
of a typical distribution identity like
\begin{align}
\Bigl[ \frac{1+z^2}{1-z} \Bigr]_+
= 2\cL_0(1-z) + \frac{3}{2}\delta(1-z) - (1+z)
\,.\end{align}
Namely, it expresses plus distributions of a function $y(z_a, z_b)$
in terms of simpler plus distributions of $1-z_{a,b}$ plus regular terms.
Moving the regular terms out of the plus distribution incurs additional boundary terms.

A key property of the left-hand side of \eq{translation_L0L0_exact} is
that it vanishes when integrated over all of $y$. It is instructive to see how
this is reproduced by the right-hand side
by projecting onto $z$ via $\int \! \df z_a \df z_b \, \delta(z - z_a z_b)$.
The only nontrivial projection integral that is required is
\begin{equation}
\int \! \df z_a \df z_b \, \delta(z - z_a z_b) \, \cL_0(1-z_a) \cL_0(1-z_b)
= 2 \cL_1(1-z) - \frac{\pi^2}{6}\delta(1-z) - \frac{\ln z}{1-z}
\,.\end{equation}
These terms precisely cancel the contributions from the other terms upon
projecting onto $z$.

\subsection{\boldmath Exact NLO partonic cross sections in terms of $(z_a, z_b)$}

Since \eq{factorization_generalized_partonic} captures
the full singularity structure as $z_a \to 1$ and/or $z_b \to 1$, the
power corrections to it are of relative $\ordsq{(1-z_a)(1-z_b)}$ and are fully integrable.
Hence, we can construct the exact partonic cross section in terms of $(z_a, z_b)$\
from the results in terms of $(z, y)$ from \refscite{Anastasiou:2003yy, Anastasiou:2002qz} as
\begin{align}
\hsigma_{ij}(z_a, z_b)
&= H_{ik} \hcI_{kj}(z_a, z_b) + H_{kj} \hcI_{ki}(z_b, z_a) - H_{ij} \hS(z_a, z_b)
\nn \\ & \quad
+ \biggl[ \frac{\df z \,\df y}{\df z_a \df z_b} \,
\hsigma_{ij}\bigl[z(z_a, z_b), y(z_a, z_b)\bigr]
- H_{ik} \hcI_{kj}(z_a, z_b) - H_{kj} \hcI_{ki}(z_b, z_a) + H_{ij} \hS(z_a, z_b) \biggr]_{z_{a,b} < 1}
\,.\end{align}
Here, the term in square brackets is evaluated in the bulk, away from any singularities,
so we can simply plug in \eq{definition_z_y}.
In the following, we collect the resulting expressions for the Drell-Yan and $gg\to H$
cross sections to NLO written explicitly in terms of $(z_a, z_b)$.

\subsubsection{Results for Drell-Yan}

The Born cross section for Drell-Yan production, $q\bar{q} \to Z/\gamma^\ast \to \ell^+ \ell^-$, is given by
\begin{align} \label{eq:born_xsec_dy}
\sigma_{B,q}^\DY
&= \frac{4\pi \alpha_\text{em}^2}{3N_c Q^2} \biggl[Q_q^2 + \frac{(v_q^2 + a_q^2) (v_\ell^2+a_\ell^2) - 2 Q_q v_q v_\ell (1-m_Z^2/Q^2)}
{(1-m_Z^2/Q^2)^2 + m_Z^2 \Gamma_Z^2/Q^4}\biggr]
\,,\end{align}
with $N_c$ the number of colors, $Q_q$ the quark charge in units of $\abs{e}$,
$v_{\ell,q}$ and $a_{\ell,q}$ the standard electroweak vector and axial couplings of the leptons and quarks,
and $m_Z$ and $\Gamma_Z$ the mass and width of the $Z$ boson.
To restrict to $q\bar{q} \to \gamma^\ast \to \ell^+ \ell^-$, only the first term $\propto Q_q^2$ is kept.
The complete LO cross section is given by
\begin{align} \label{eq:LO_xsec_dy}
\frac{\df \sigma^\DY_\text{LO}}{\df Q \, \df Y}
= \frac{2Q}{\Ecm^2} \, \frac{\df \sigma^\DY_\text{LO}}{\df x_a \df x_b}
= \frac{2Q}{\Ecm^2} \, \sum_q \sigma_{B,q}^\DY \, \bigl[ f_q(x_a) \, f_\bq(x_b) + f_\bq(x_a) \, f_q(x_b) \bigr]
\,,\end{align}
where the sum runs over $q = \{ u,d,c,s,b \}$.
We expand the partonic cross section for Drell-Yan as
\begin{align} \label{eq:partonic_xsec_dy_expansion}
\hat{\sigma}_{ij}(z_a,z_b,Q,\mu) = \sum_{n=0}^\infty \Bigl[\frac{\as(\mu)}{4\pi}\Bigr]^n\, \sigma_{ij}^{(n)}(z_a,z_b,Q,\mu)
\,.\end{align}
The LO result corresponding to \eq{LO_xsec_dy} is given by $\hsigma^{(0)}_{q\bq}(z_a, z_b, Q, \mu) = \sigma_{B,q}^\DY\, \delta(1-z_a)\,\delta(1-z_b)$.
Writing $\bz_a \equiv 1 - z_a$ and $\bz_b \equiv 1 - z_b$ for short, the NLO coefficient for the $q\bq$ channel is given by
\begin{align} \label{eq:partonic_xsec_dy_qqbar}
\frac{1}{\sigma_{B,q}^\DY} \frac{\hsigma^{(1)}_{q\bq}(z_a, z_b, Q, \mu)}{C_F} &= \delta(\bz_a) \delta(\bz_b) (2\pi^2 - 16) + 4\cL_1(\bz_a) \delta(\bz_b) + 4\cL_0(\bz_a) \cL_0(\bz_b) + 4 \delta(\bz_a) \cL_1(\bz_b)
\nn \\ & \quad + \biggl\{
   - 2(1 + z_b) \cL_0(\bz_a)
   + \delta(\bz_a) \Bigl[ 2\bz_b - 4(1 + z_b) \ln \bz_b - \frac{2(1+z_b^2) \ln z_b}{\bz_b}
   \nn \\ & \qquad \quad
   + \frac{4}{\bz_b}\ln \frac{2z_b}{1+z_b} + 2(1+z_b) \ln \frac{1-z_b^2}{2z_b}
   \Bigr]
    - 4 \ln \frac{\mu}{Q} \, \delta(\bz_a) \Bigl[ 2\cL_0(\bz_b) + \frac{3}{2} \delta(\bz_b) - (1+z_b)  \Bigr]
    \nn \\ & \qquad \quad
  + \frac{
   2(z_a^2 + z_b^2)[(1+z_a)^2 + z_az_b(3+2z_a+z_az_b)]
}{
(1+z_a)(1+z_b)(z_a + z_b)^2
}
+ (z_a \leftrightarrow z_b)
\biggr\}
\,,\end{align}
where $(z_a \leftrightarrow z_b)$ indicates all previous expressions in the curly brackets repeated with $z_a$ and $z_b$ interchanged.
For the $qg$ channel we have
\begin{align} \label{eq:partonic_xsec_dy_qg}
\frac{1}{\sigma_{B,q}^\DY} \frac{\hsigma_{qg}^{(1)}(z_a, z_b, Q, \mu)}{T_F} &= 2( z_b^2 + \bz_b^2) \cL_0(\bz_a) + \delta(\bz_a) \Bigl[ 2(z_b^2 + \bz_b^2) \ln \frac{2\bz_b}{1+z_b} + 4 z_b \bz_b \Bigr]
- 4 \ln \frac{\mu}{Q} \, \delta(\bz_a) \bigl( z_b^2 + \bz_b^2 \bigr)
\\ & \quad
+ \frac{1}{(1 +z_a)(z_a + z_b)^3} \Bigl[
   -4 z_a^5 z_b^3
   -4 z_a^4 z_b^2 (-1 + z_b + 2 z_b^2) 
   +2z_a^3 (1 + 4 z_b^2 + 2 z_b^3 - 4 z_b^4 - 4 z_b^5)
   \nn \\ &\qquad
   +2 z_a^2 z_b (1 + 4 z_b + 8 z_b^2 - 8 z_b^3 - 4 z_b^4)
   +2 z_a z_b^2 (1 + 4 z_b - 2 z_b^2 - 4 z_b^3)
   -2 z_b^3 (1 - 2 z_b + 2 z_b^2) \Bigr]
\nn \,.\end{align}
The $gq$ channel is given by $\sigma_{gq}^{(1)}(Q, \mu, z_a, z_b) = \sigma_{qg}^{(1)}(Q, \mu, z_b, z_a)$.
The results for $q \leftrightarrow \bq$ are identical.

\subsubsection{Results for gluon-fusion Higgs production}

For gluon-fusion Higgs production, $gg \to H$, we use the effective Lagrangian
in the limit $m_H^2 \ll 4 m_t^2$,
\begin{equation} \label{eq:def_C_t}
\mathcal{L}_\mathrm{eff}(m_H) = - \frac{C_t}{12\pi v}\,\as\, G_{\mu\nu}^a G^{a,\mu\nu} H
\,, \qquad
C_t = 1 + \frac{\as}{4\pi} (5 C_A - 3 C_F) + \ord{\as^2}
\,.\end{equation}
As in \refcite{Anastasiou:2002qz}, the Wilson coefficient $C_t$ is always perturbatively expanded
against other fixed-order contributions.
The Born cross section and LO rapidity spectrum are given by
\begin{align} \label{eq:born_xsec_ggH}
\sigma_B^\ggH = \frac{1}{72\pi v^2 (N_c^2 - 1)}
\,, \qquad
\frac{\df \sigma^\ggH_\text{LO}}{\df Y}
= x_a x_b \, \sigma_B^\ggH \, \as^2 \,f_g(x_a) \, f_g(x_b)
\,.\end{align}
We write the partonic cross section as
\begin{align} \label{eq:partonic_xsec_ggH_expansion}
\hat{\sigma}_{ij}(z_a,z_b,m_t,m_H,\mu) \equiv \sigma_B^\ggH\, \as^2(\mu)\, |C_t(m_t, \mu)|^2\, \sum_{n=0}^\infty \Bigl[\frac{\as(\mu)}{4\pi}\Bigr]^n\, \heta_{ij}^{(n)}(z_a,z_b,m_H,\mu)
\,.\end{align}
The NLO coefficient function for the $gg$ channel is given by
\begin{align}  \label{eq:partonic_xsec_ggH_gg}
\frac{\heta_{gg}^{(1)}(z_a, z_b,m_H,\mu)}{C_A}
&= 2\pi^2 \delta(\bz_a) \delta(\bz_b) + 4 \cL_1(\bz_a) \delta(\bz_b) + 4 \delta(\bz_a) \cL_1(\bz_b) + 4 \cL_0(\bz_a) \cL_0(\bz_b)
\\ &\quad
+ \biggl\{
4 \cL_0(\bz_a) \Bigl[ \frac{1}{z_b} - 2 + z_b - z_b^2 \Bigr]
+ 4\delta(\bz_a) \frac{1}{\bz_b} \Bigl[
     \frac{\ln(2\bz_b)}{z_b}
   - 3\ln \bz_b
   + 2\ln \frac{1+z_b}{2}
   - \frac{\ln(1+z_b)}{z_b}
   \nn \\ & \qquad \quad
   - z_b(3-2z_b + z_b^2) \ln \frac{1+z_b}{2\bz_b}
\Bigr]
 - 4\ln\frac{\mu}{m_H} \delta(\bz_a) \Bigl[ 2\mathcal{L}_0(\bz_b) \frac{(1-z_b + z_b^2)^2}{z_b} \Bigr]
\nn \\ & \qquad \quad
+ \frac{4z_b}{z_a (1+z_a) (1+z_b) (z_a+z_b)^4}\Bigl[2 z_b^2 + z_b^3 + 3 z_a^6 z_b^4 + 2 z_a^5 z_b^3 (5 + 5 z_b + 2 z_b^2)
\nn \\ & \qquad \qquad
+ z_a^4 z_b^2 (16 + 17 z_b + 12 z_b^2 + 6 z_b^3 + 2 z_b^4)
 + z_a^3 z_b (5 + 22 z_b + 12 z_b^2 + 8 z_b^3 + 8 z_b^4 + 2 z_b^5)
\nn \\ & \qquad \qquad
+ z_a^2 (3 + 2 z_b^2 + 7 z_b^3 + 2 z_b^4 + 4 z_b^5 + 2 z_b^6)
+ z_a z_b (4 + z_b + z_b^2 + z_b^3 + z_b^5)
\Bigr]
+ (z_a \leftrightarrow z_b)
 \biggr\}
\nn\,.\end{align}
For the $gq$ and $qg$ channels we find, with $\eta_{qg}^{(1)}(z_a, z_b, m_H, \mu) = \eta_{gq}^{(1)}(z_b, z_a, m_H, \mu)$,
\begin{align} \label{eq:partonic_xsec_ggH_gq}
\frac{\heta_{gq}^{(1)}(z_a, z_b, m_H, \mu)}{C_F}
&= 2 \cL_0(\bz_a) \, \frac{2-2z_b + z_b^2}{z_b}
+ 2 \delta(\bz_a) \Bigl[ z_b + \frac{2-2z_b + z_b^2}{z_b} \Bigl( \ln \frac{2\bz_b}{1+z_b} - 2\ln \frac{\mu}{m_H} \Bigr) \Bigr]
\nn \\ & \quad
+ \frac{2}{(1+z_a)z_b(z_a + z_b)^3} \Bigl[
   z_a^3 (2 - 2 z_b + z_b^2) + z_a^2 (4 - 2 z_b - 4 z_b^2 + 7 z_b^3 - 2 z_b^5) 
   \nn \\ &\qquad
   + z_a z_b (4 - 4 z_b + 4 z_b^2 + z_b^3 - 2 z_b^4) - z_b^2 (-2 + 2 z_b - 2 z_b^2 + z_b^3) \Bigr]
\,.\end{align}
The $q\bar{q}$ channel is fully regular and given by
\begin{align} \label{eq:partonic_xsec_ggH_qqbar}
\frac{\heta_{q\bar{q}}^{(1)}(z_a, z_b, m_H, \mu)}{C_F}
= \frac{N_c^2 - 1}{N_c} \, \frac{4(1 + z_a z_b) (z_a^4 z_b^2+z_a^2 z_b^4-4 z_a^2 z_b^2+z_a^2+z_b^2)}{(z_a+z_b)^4}
\,.\end{align}
The $N_c$-dependent prefactor accounts for the different color average compared
to the Born cross section.

The above expressions are in full agreement with \refscite{Mathews:2004xp, Ravindran:2006bu},
which in turn agree with the earliest result for the NLO Drell-Yan rapidity spectrum in \refcite{Kubar:1980zv}.
In \refscite{Mathews:2004xp, Ravindran:2006bu},
the cross section was also parametrized in terms of $x_{a,b}$ and $z_{a,b}$,
but all subtractions were written out in full at the level of the hadronic cross section.

In \refcite{Anastasiou:2003yy}, only the sum of the $gq$ and $qg$ coefficient functions for
Higgs productions was given. The separation of the singular terms into the two channels is unique
because only the $gq$ ($qg$) channel can be singular as $y \to 0$ ($y \to 1$).
We determined the separation of the regular terms by comparing to \refcite{Ravindran:2006bu}.
To the best of our knowledge, this is the first time that the explicit analytic agreement between
the independent NLO calculations in terms of $(z_a, z_b)$ and $(z,y)$ has been established.

Finally, we have also compared our numerical implementation of \eq{partonic_xsec_dy_expansion}-\eqref{eq:partonic_xsec_ggH_qqbar}
with the rapidity spectra obtained from \texttt{Vrap~0.9}~\cite{Anastasiou:2003ds} for Drell-Yan and from \texttt{SusHi~1.7.0}~\cite{Harlander:2012pb, Harlander:2016hcx} for Higgs production,
finding excellent agreement. Since \texttt{Vrap 0.9} implements the $(z,y)$ parametrization,
this effectively confirms the distributional identities in \tab{distribution_dictionary_z_y}
numerically with physical PDFs as test functions.

\subsection{Comment on rapidity-dependent soft threshold results in the literature}

As we have discussed, the soft threshold limit is fully contained in the
generalized threshold limit. Our results thus provide an independent
confirmation that \eq{factorization_soft} is the correct soft threshold
factorization for the cross section differential in both $Q$ and $Y$, or
equivalently \eq{factorization_soft_partonic} for the two-dimensional partonic
cross section in $(z_a, z_b)$.

Several results in the literature~\cite{Bolzoni:2006ky, Mukherjee:2006uu,
Becher:2007ty, Bonvini:2010tp, Bonvini:2015ira} considering the soft threshold
factorization differential in rapidity differ from \eq{factorization_soft}. The
difference is manifest already at fixed NLO in the term $\cL_0(1-z)[\cL_0(y) +
\cL_0(1-y)]$, which appears in the flavor-diagonal partonic cross sections. The
distributional identity in \eq{translation_L0L0_exact} unambiguously shows that
this term has a double singularity in the limit $z_a \to 1$ and $z_b \to 1$,
which means it contributes a priori at leading power in the soft limit $z = z_a
z_b \to 1$, i.e., it contributes to the $m_a = m_b = -1$ term in
\eq{power_expansion_general}. This can already be seen just by considering the
distribution in the bulk since
\begin{equation}
\df z \, \df y \, \frac{1}{1-z}\Bigl(\frac{1}{y} + \frac{1}{1-y}\Bigr)
= \df z_a \df z_b \, \frac{1}{(1-z_a)(1-z_b)}\bigl[1 + \ord{1-z_a, 1-z_b}\bigr]
\,.\end{equation}
Moreover, \eq{translation_L0L0_exact} shows that it contributes at
leading-logarithmic order.
The soft function in \eqs{factorization_soft}{factorization_soft_partonic}
precisely contains the leading-power contribution of this term, which is given
by the first four terms on the right-hand side of \eq{translation_L0L0_exact}.

By contrast, this term and analogous ones at higher order are missing in the
leading-power resummed results in \refscite{Bolzoni:2006ky, Mukherjee:2006uu, Becher:2007ty, Bonvini:2010tp, Bonvini:2015ira}.
There, it is effectively argued that the contribution of such terms to the rapidity spectrum
is power-suppressed in $1-z$, leading to the incorrect conclusion
that the rapidity dependence in the soft threshold limit can be included
simply by taking $\hsigma_{ij}(z_a, z_b)$ to be $\hsigma_{ij}(z)\, [\delta(y) + \delta(1-y)]/2$
or, depending on the reference, $\hsigma_{ij}(z) \, \delta(y - 1/2)$, where $\hsigma_{ij}(z)$
is the inclusive, rapidity-integrated, partonic cross section.
In the following, we give a critical appraisal of the arguments used to support
this conclusion and show why they are flawed.

This replacement first appeared in \refcite{Laenen:1992ey}, where it was conjectured
to provide an approximation to the threshold-resummed rapidity spectrum at small $Y$.
The phenomenological impact of the correct convolution structure on PDF
determinations relying on soft threshold resummation was discussed in
\refcite{Westmark:2017uig}. A detailed numerical study of the difference at the
level of the resummed Drell-Yan rapidity spectrum was performed in
\refcite{Banerjee:2018vvb}.

\paragraph{Argument based on PDF momentum fractions}

What makes the $\cL_0(1-z)[\cL_0(y) + \cL_0(1-y)]$ term subtle is that it vanishes
upon integration over $y$, so it drops out in the inclusive cross section. This
fact alone is of course insufficient to argue that it is power suppressed at each point
in the spectrum. It simply means that different leading-power terms conspire to
cancel upon integration, which is clear in terms of $(z_a, z_b)$, as discussed
below \eq{translation_L0L0_exact}.

The argument in \refcite{Becher:2007ty} rests on the observation that the PDF arguments,
$x_a/z_a(z, y)$ and $x_b/z_b(z, y)$, in the two-dimensional convolution integral are
independent of $y$ at $z = 1$, from
which it is concluded that the $y$ dependence of the PDF arguments is power suppressed
in $1-z$ and can be dropped. If this is done, the $y$ integral
becomes unconstrained and can be carried out freely, which eliminates this term.
More generally, one could then replace
$\hsigma_{ij}(z, y) = \hsigma_{ij}(z) \, \delta(y-1/2)[1 + \ord{1-z}]$
underneath the convolution integral.

However, a closer inspection of the PDF arguments in terms of $(z, y)$ reveals that
\begin{equation}
\frac{x_a}{z_a(z,y)}
= x_a\Bigl\{1 + y(1-z) + \ordsq{(1-z)^2}\Bigr\}
\,, \qquad
\frac{x_b}{z_b(z,y)}
= x_b\Bigl\{1 + (1-y)(1-z) + \ordsq{(1-z)^2}\Bigr\}
\,.\end{equation}
Hence, the $y$ dependence is not power suppressed but multiplies the
\emph{leading} dependence of the PDF arguments on $z$ itself, and so it cannot
be dropped. This becomes even clearer in terms of $(z_a, z_b)$,
\begin{equation}
y(1-z) = 1-z_a + \ordsq{(1-z)^2}
\,, \qquad
(1-y)(1-z) = 1-z_b + \ordsq{(1-z)^2}
\,,\end{equation}
which are precisely the $\ord{\lambda^2}$ convolution variables $k^\mp/(Qe^{\pm Y})$
in \eq{factorization_soft}.

To illustrate explicitly that the $y$ dependence has a leading-power effect,
consider the hadronic soft threshold limit $1-x_a \sim 1-x_b \ll 1$
and a simple toy PDF with a power-law behavior near the endpoint, with $\alpha > 0$,
\begin{equation}
f(x) \equiv \theta(1-x) (1-x)^\alpha
\,.\end{equation}
Using \eq{translation_L0L0_exact}, it is straightforward to show that $\cL_0(1-z)[\cL_0(y) + \cL_0(1-y)]$ gives
rise to double logarithms of $1-x_{a,b}$, but performing the integral directly
in terms of $(z,y)$ is tedious, essentially as tedious as deriving
\eq{translation_L0L0_exact} itself. Instead, to disprove the above argument
and show that the $y$ dependence is not power suppressed, it suffices to consider
two terms that have the same $y$ integral,
\begin{align} \label{eq:toy_example_def_a_b}
A(z,y)
\equiv \cL_0(1-z)\, \frac{\delta(y) + \delta(1-y)}{2}
\,, \qquad
B(z,y) \equiv \cL_0(1-z) \, \delta\Bigl( y - \frac{1}{2} \Bigr)
\,,\end{align}
and show that they give different results at leading power, while the above argument
would imply that they do not.
Convolving $A(z, y)$ and $B(z, y)$ against the toy PDFs over the domain shown in \fig{integration_region} yields
\begin{align}
\int \! \df z \, \df y \, A(z,y) \, f\Bigl[ \frac{x_a}{z_a(z,y)} \Bigr] f\Bigl[ \frac{x_b}{z_b(z,y)} \Bigr]
&= f(x_a) f(x_b) \Bigl[ \frac{1}{2}\ln \bigl(1-x_a\bigr) + \frac{1}{2}\ln \bigl(1-x_b\bigr) - H_\alpha + \ord{1\!-\!x_a, 1\!-\!x_b} \Bigr]
\,, \nn \\
\int \! \df z \, \df y \, B(z,y) \, f\Bigl[ \frac{x_a}{z_a(z,y)} \Bigr] f\Bigl[ \frac{x_b}{z_b(z,y)} \Bigr]
&= f(x_a) f(x_b) \Bigl[ \ln \bigl(1 -\max \{ x_a, x_b \}\bigr) - H_\alpha + \ord{1\!-\!x_a, 1\!-\!x_b} \Bigr]
\,,\end{align}
where $H_\alpha$ is the harmonic number.
To evaluate the integrals it is convenient to already expand at integrand level,
e.g. $1-x_a/z_a = 1-x_a + y(1-z)$ up to higher powers in $1-z$ and $1-x_a$.
The maximum in the second case arises because the integration region in $z$ along
fixed $y = 1/2$ is cut off by the square of the larger of the two momentum fractions.
(The order of expanding in $1-x_a$ and $1-x_b$ also needs to be picked accordingly.)
Clearly, the two results only coincide for $Y = 0$, where $x_a = x_b$.
Away from $Y = 0$, the logarithmic dependence on $x_{a,b}$ and thus on $Y$ differs at leading power.

\paragraph{Fourier-transform argument}

An alternative line of argument~\cite{Bolzoni:2006ky, Mukherjee:2006uu, Bonvini:2010tp}
relies on taking the Fourier transform of the partonic cross section
to also argue that the $y$ dependence is trivial.
A first step is to change variables from $y$ to
\begin{equation}
u \equiv \frac{1}{2} \ln \frac{z_a}{z_b}
\,, \qquad -\umax \leq u \leq \umax
\,, \qquad \umax \equiv \ln \frac{1}{\sqrt{z}}
\,.\end{equation}
(Note that the variables $u$ and $y$ are precisely interchanged in the notation used in \refcite{Bonvini:2010tp}.)
One then considers the Fourier transform of the partonic cross section $C(z,u)$ with respect to $u$,
\begin{equation}
\tilde{C}(z,M) \equiv \int \! \df u \, e^{\img M u} \, C(z, u)
\stackrel{?}{=} \int \! \df u \, C(z, u) \bigl[ 1 + \ord{1-z}\bigr]
\,.\end{equation}
The second equality, which is in question, is based on observing that $C(z, u)$ only has support on an
interval bounded by $\umax \sim 1-z$, and concluding that the Fourier kernel can be expanded
in $u\sim 1-z$ as $e^{\img M u} \stackrel{?}{=} 1 + \ord{1-z}$, and so $\tilde{C}(z,M)$ is independent of $M$
at leading power. Thus, taking the inverse Fourier transform, the partonic cross section may be approximated as
\begin{equation} \label{eq:every_function_is_trivial}
C(z,u) = \int \! \frac{\df M}{2\pi} \, e^{-\img M u} \, \tilde{C}(z,M) \stackrel{?}{=} \delta(u) \int \! \df u' \, C(z, u') \bigl[ 1 + \ord{1-z}\bigr]
\,.\end{equation}
This argument is flawed because in order to satisfy the
Fourier inversion theorem, one must count $M \sim (1-z)^{-1}$ if one wants to
count $u \sim 1-z$. In particular, one is not allowed to count $M \sim 1$ when taking the
limit $z \to 1$ (or equivalently $N \to \infty$ for the Mellin conjugate $N$ of
$z$). This is essential because $C(z,u)$ contains distributional terms in $u$
that cancel the suppression by the integration domain.

To disprove \eq{every_function_is_trivial}, it again suffices to consider $A(z, y)$ and $B(z, y)$
defined in \eq{toy_example_def_a_b}. Changing variables to $u$, we have
\begin{equation}
\df y \, A(z,y)
= \df u \, \cL_0(1-z)\, \frac{\delta(u + \umax) + \delta(u - \umax)}{2}
\,, \qquad
\df y \, B(z,y)
= \df u \, \cL_0(1-z) \, \delta( u )
\,.\end{equation}
Both terms satisfy the assumptions of the above argument, i.e., they only have support for $\abs{u} \leq \umax$.
Changing variables back to $y$, \eq{every_function_is_trivial} would
imply that up to power corrections in $1-z$,
\begin{equation}
A(z,y) = \cL_0(1-z)\, \frac{\delta(y) + \delta(1-y)}{2}
\stackrel{?}{=} \cL_0(1-z) \, \delta\Bigl( y - \frac{1}{2} \Bigr)
= B(z,y)
\,.\end{equation}
In fact, the overall factor found in \refcite{Bonvini:2010tp} is $\delta(y-1/2)$,
while it is $[\delta(y) + \delta(1-y)]/2$ in \refcite{Becher:2007ty},
and the above argument was used in \refcite{Bonvini:2010tp} to argue that the two
are equivalent. As a distributional identity, this is obviously incorrect.
The only thing that is equal between $A(z, y)$ and $B(z, y)$ are their $y$ integrals,
and as demonstrated before, this is insufficient because the $y$ dependence of the PDF
arguments is a leading-power effect and cannot be neglected.

\paragraph{Summary}

For $pp$ production processes in general,
to correctly describe the soft threshold limit of differential observables
that are sensitive to the total rapidity of the Born system, one must maintain the
two-dimensional dependence on $z_a$ and $z_b$ in the convolutions against the PDFs.
Equivalently, in Mellin space one must maintain two Mellin fractions
$N_a$ and $N_b$ as in the original \refcite{Catani:1989ne}.
In terms of the Mellin conjugate $N$ of $z$ and a Fourier conjugate $M$ of another
variable like $u$, one has to keep the dependence on $M \sim \abs{N_a - N_b} \sim N$.
In particular, reducing the two-dimensional convolution structure
to one dimension -- such that the rapidity dependence is only carried by
the luminosity function -- amounts to making an additional assumption that is not
justified by taking the soft limit.

\subsection{Perturbative ingredients}

Here, we collect the required perturbative ingredients to evaluate the generalized threshold
factorization theorem to two loops as well as the highest few logarithmic terms at three loops.

\subsubsection{Hard functions}

The hard function for Drell-Yan production, $q\bar{q} \to Z/\gamma^\ast \to \ell^+ \ell^-$, is given by
\begin{align} \label{eq:hard_function_dy}
H_{ij}^\DY(Q^2, \mu)
&= \sum_q \sigma_{B,q}^\DY \, \bigl(\delta_{iq}\delta_{j\bq} + \delta_{i\bq}\delta_{jq}\bigr) \, \abs{C_{q\bq}^{V}(Q^2, \mu)}^2
\,,\end{align}
where the sum runs over $q = \{ u,d,c,s,b \}$,
$\sigma_{B,q}^\DY$ is the Born cross section given in \eq{born_xsec_dy},
and $C_{q\bq}^{V}$ is the Wilson coefficient from matching the QCD quark vector current onto SCET.
In principle, we also need the matching coefficient $C_{q\bq}^{A}$ for the axial-vector current,
which differs from $C_{q\bq}^{V}$ starting at $\ord{\as^2}$ by small singlet corrections due to the
large mass splitting between bottom and top quarks.
We neglect these terms, as is done in \texttt{Vrap 0.9}, and use $H_{ij}^\DY \propto |C_{q\bq}^{V}|^2$ throughout.
The hard function for gluon-fusion Higgs production in the limit $m_H^2 \ll 4m_t^2$ reads
\begin{align} \label{eq:hard_function_ggH}
H_{ij}^\ggH(m_t, m_H, \mu)
&= \sigma_B^\ggH \,\bigl\vert \as(\mu) \, C_t(m_t, \mu)\bigr\vert^2 \, \delta_{ig}\delta_{jg} \, \abs{C_{gg}(m_H, \mu)}^2
\,,\end{align}
where $\sigma_B^\ggH$ and $C_t$ are given in \eqs{def_C_t}{born_xsec_ggH}.
The Wilson coefficient $C_{gg}$ arises from matching the $gg\to H$ operator in \eq{def_C_t} onto SCET.
The Wilson coefficients contain the IR-finite virtual corrections to the Born process. They are normalized
as $C = 1 + \ord{\as}$ and can be found in \refcite{Ebert:2017uel} in our notation.

In the main text, we also refer to the fixed-order expansion of the hard function,
where the coefficients $H_{ij}^{(n)}$ include all prefactors that are present at Born level,
\begin{equation}
H_{ij}(Q^2, \mu) = \sum_{n=0}^\infty \Bigl[ \frac{\as(\mu)}{4\pi} \Bigr]^n H_{ij}^{(n)}(Q^2, \mu)
\,.\end{equation}

\subsubsection{Beam functions}

For $t \gg \LQCD^2$, the modified beam function defined by \eq{def_modified_beam_function}
can be matched onto PDFs as
\begin{align} \label{eq:modified_beam_function_matching}
\tB_i(t, x, \mu) = \int \! \frac{\df z}{z} \, \tcI_{ij}(t, z, \mu) \, f_j\Bigl( \frac{x}{z}, \mu \Bigr) \,
\Bigl[ 1 + \ORd{\frac{\LQCD^2}{t}} \Bigr]
\,,\end{align}
which directly follows from the analogous matching relations for the
inclusive and double-differential beam functions, $B_i(t, x, \mu)$ and $B_i(t, \vec{k}_T, x, \mu)$,
with matching coefficients $\cI_{ij}(t, z, \mu)$ and $\cI_{ij}(t, \vec{k}_T, z, \mu)$~\cite{Stewart:2009yx, Stewart:2010qs, Jain:2011iu}.

The modified beam function satisfies the same RGE as the other beam functions,
\begin{align} \label{eq:double_differential_beam_function_rge}
\mu \frac{\df}{\df \mu} \tB_i(t, x, \mu)
= \int \! \df t' \, \gamma_B^i(t-t', \mu) \, \tB_i(t', x, \mu)
\,, \quad
\gamma_B^i(t, \mu) = -2 \Gamma_\cusp^i\bigl[ \as(\mu) \bigr] \, \cL_0(t, \mu^2) + \gamma_B^i\bigl[ \as(\mu) \bigr] \delta(t)
\,,\end{align}
because the RGE does not change the $\vec{k}_T$ dependence~\cite{Jain:2011iu}.
The matching coefficients satisfy the RGE~\cite{Stewart:2010qs}
\begin{align} \label{eq:modified_matching_coefficient_rge}
\mu \frac{\df}{\df \mu} \tcI_{ij}(t, z, \mu)
&= \int \! \df t' \, \int \! \frac{\df z'}{z'} \, \tcI_{ik}\Bigl(t - t', \frac{z}{z'}, \mu\Bigr) \, \Bigl\{ \delta_{kj} \delta(1-z') \, \gamma_B^i(t', \mu) - 2\delta(t')\,P_{kj}\bigl[\as(\mu), z' \bigr] \Bigr\}
\,,\end{align}
where $P_{ij}(\as, z)$ is the PDF anomalous dimension.
We define the perturbative expansion of $\tcI_{ij}$ as
\begin{align} \label{eq:modified_beam_function_pert_series}
\tcI_{ij}(t, z, \mu) = \sum_{n=0}^\infty \Bigl[ \frac{\as(\mu)}{4\pi} \Bigr]^n \tcI_{ij}^{(n)}(t, z, \mu)
\,.\end{align}
Solving \eq{modified_matching_coefficient_rge} order by order in $\as$, we obtain a
general expression for $\tcI_{ij}^{(n)}(t, z, \mu)$
\begin{align} \label{eq:modified_beam_function_master_formula}
\tcI_{ij}^{(0)}(t, z,\mu)
&= \delta(t) \, \delta_{ij}\delta(1-z)
\,, \nn \\
\tcI_{ij}^{(1)}(t, z,\mu)
&= \mathcal{L}_1(t,  \mu^2)\, \Gamma_0^i \, \delta_{ij}\delta(1-z)
+ \mathcal{L}_0(t,  \mu^2) \Bigl[ P^{(0)}_{ij}(z) - \frac{\gamma_{B\,0}^i}{2}\, \delta_{ij} \delta(1-z) \Bigr]
+ \delta(t) \, \tI^{(1)}_{ij}(z)
\,, \nn \\
\tcI_{ij}^{(2)}(t, z,\mu)
&= \mathcal{L}_3(t,  \mu^2)\,  \frac{(\Gamma_0^i)^2}{2} \, \delta_{ij} \delta(1-z)
\nn \\ &\quad
+ \mathcal{L}_2(t,  \mu^2) \, \frac{\Gamma_0^i}{2} \biggl[
   -\Bigl(\beta_0  + \frac{3}{2}\gamma_{B\,0}^i\Bigr)\delta_{ij}\delta(1-z)  + 3 P^{(0)}_{ij}(z)
\biggr]
\nn \\ &\quad
+ \mathcal{L}_1(t,  \mu^2) \biggl[
   \Bigl(
   - \frac{\pi^2}{6} (\Gamma_0^i)^2 + \frac{\beta_0}{2}\gamma_{B\,0}^i + \frac{(\gamma_{B\,0}^i)^2}{4} + \Gamma_1^i
   \Bigr)\delta_{ij} \delta(1-z)
   - \bigl(\beta_0 + \gamma_{B\,0}^i\bigr) P^{(0)}_{ij}(z)
   \nn \\ &\qquad\qquad\qquad
   + \bigl(P^{(0)}_{ik} \otimes P^{(0)}_{kj}\bigr)(z)
   + \Gamma_0^i \tI^{(1)}_{ij}(z)
\biggr]
\nn \\ &\quad
+ \mathcal{L}_0(t,  \mu^2) \biggl[
   \Bigl(
      \zeta_3 (\Gamma_0^i)^2 + \frac{\pi^2}{12} \Gamma_0^i \gamma_{B\,0}^i - \frac{\gamma_{B\,1}^i}{2}
   \Bigr) \delta_{ij} \delta(1-z)
   - \frac{\pi^2}{6} \Gamma_0^i P^{(0)}_{ij}(z)
   + P^{(1)}_{ij}(z)
   \nn \\ &\qquad\qquad\qquad
   - \Bigl(\beta_0  + \frac{\gamma_{B\,0}^i}{2} \Bigr) \tI^{(1)}_{ij}(z)
   + \bigl(\tI^{(1)}_{ik} \otimes P^{(0)}_{kj}\bigr)(z)
\biggr]
\nn \\ &\quad
+ \delta(t) \, \tI^{(2)}_{ij}(z)
\,, \nn \\
\tcI_{ij}^{(3)}(t, z,\mu)
&= \mathcal{L}_5(t,  \mu^2)\, \frac{(\Gamma_0^i)^3 }{8} \delta_{ij} \delta(1-z)
\nn \\ &\quad
+ \mathcal{L}_4(t, \mu^2) \, \frac{5}{8}(\Gamma_0^i)^2 \biggl[
   - \Bigl(\frac{2}{3} \beta_0 + \frac{\gamma_{B\,0}^i}{2} \Bigr)\delta_{ij} \delta(1-z)
   + P^{(0)}_{ij}(z)
\biggr]
\nn \\ &\quad
+ \mathcal{L}_3(t, \mu^2) \, \Gamma_0^i \biggl[
   \Bigl( -\frac{\pi^2}{6} (\Gamma_0^i)^2 + \frac{\beta_0^2}{3}
   + \frac{5}{6}\beta_0 \gamma_{B\,0}^i + \frac{(\gamma_{B\,0}^i)^2}{4} + \Gamma_1^i
   \Bigr) \delta_{ij} \delta(1-z)
   - \Bigl(\frac{5}{3}\beta_0 + \gamma_{B\,0}^i \Bigr)P^{(0)}_{ij}(z)
   \nn \\ &\qquad\qquad\qquad\quad
   + \bigl(P^{(0)}_{ik} \otimes P^{(0)}_{kj}\bigr)(z)
   + \frac{\Gamma_0^i }{2} \, \tI^{(1)}_{ij}(z)
\biggr]
\nn \\ &\quad
+ \dotsb + \delta(t) \, \tI^{(3)}_{ij}(z)
\,,\end{align}
where $\cL_n(t, \mu^2) \equiv (1/\mu^2)\cL_n(t/\mu^2)$,
$\tI^{(n)}_{ij}(z)$ is the $\ord{\as^n}$ boundary term that is not predicted by the RGE,
and the ellipses in the three-loop result indicate the terms proportional to $\cL_{0,1,2}(t, \mu^2)$.
We also introduced the shorthand $(g \otimes h)(z) \equiv \int_z^1 \! \df z'/z' \, g(z') \, h(z/z')$
for the Mellin convolution of two functions of $z$.
In practice, we use the \texttt{MT} package~\cite{Hoeschele:2013gga}
to evaluate them analytically.
Expanding the three-loop expression for $\tcI_{ij}(t, z, \mu)$
against the hard function yields \eq{partonic_xsec_three_loops} in the main text.

The anomalous dimension coefficients in \eq{modified_beam_function_master_formula} are
as follows. The QCD $\beta$ function coefficients are
\begin{align}
\beta(\as) = -2\as \sum_{n = 0}^\infty \beta_n \left( \frac{\as}{4\pi} \right)^{n+1}
\,, \qquad
\beta_0 = \frac{11}{3}\,C_A -\frac{4}{3}\,T_F\,n_f
\,, \qquad
\beta_1
= \frac{34}{3}\,C_A^2 - 2T_F\,n_f \Bigl(\frac{10}{3}\, C_A + 2 C_F\Bigr)
\,.\end{align}
The anomalous dimensions are expanded as
\begin{equation}
\Gamma^i_\cusp(\as) = \sum_{n = 0}^\infty \Gamma^i_{n} \Bigl( \frac{\as}{4\pi} \Bigr)^{n+1}
\,,\qquad
\gamma_B^i(\as) = \sum_{n = 0}^\infty \gamma^i_{B\,n} \Bigl( \frac{\as}{4\pi} \Bigr)^{n+1}
\,,\qquad
P_{ij}(\as, z)
= \sum_{n = 0}^\infty P^{(n)}_{ij}(z) \left( \frac{\as}{4\pi} \right)^{n + 1}
\,.\end{equation}
The coefficients of the cusp anomalous dimension to two loops are~\cite{Korchemsky:1987wg}
\begin{align}
\Gamma_0^i = 4C_i
\,,\nn \qquad
\Gamma_1^i
= 4 C_i \Bigl[ C_A \Bigl( \frac{67}{9} - \frac{\pi^2}{3} \Bigr)  - \frac{20}{9}\,T_F\, n_f \Bigr]
= \frac{4}{3} C_i \bigl[ (4 - \pi^2) C_A + 5 \beta_0 \bigr]
\,,\end{align}
where $C_q = C_F$ and $C_g = C_A$.
The beam function anomalous dimension coefficients are~\cite{Stewart:2010qs, Berger:2010xi, Gaunt:2014xga, Gaunt:2014cfa}
\begin{alignat}{3}
\gamma_{B\,0}^q
&= 6C_F
\,, \qquad&
\gamma_{B\,1}^q
&= C_F \Bigl[ \Bigl(\frac{146}{9} - 80 \zeta_3\Bigr) C_A
+ (3 - 4 \pi^2 + 48 \zeta_3) C_F
+ \Bigl(\frac{121}{9} + \frac{2\pi^2}{3} \Bigr) \beta_0 \Bigr]
\,, \nn \\
\gamma_{B\,0}^g
&= 2 \beta_0
\,, \qquad&
\gamma_{B\,1}^g
&= \Bigl(\frac{182}{9} - 32\zeta_3\Bigr)C_A^2 +
\Bigl(\frac{94}{9}-\frac{2\pi^2}{3}\Bigr) C_A\, \beta_0 + 2\beta_1
.\end{alignat}
The one-loop PDF anomalous dimensions are
\begin{align}
P^{(0)}_{q_iq_j}(z) &= P^{(0)}_{\bar q_i \bar q_j}(z) = 2C_F \,\delta_{ij} \, \theta(z) P_{qq}(z)
\,,\quad
&P^{(0)}_{q_ig}(z) &= P^{(0)}_{\bar q_i g}(z) = 2T_F \,\theta(z) P_{qg}(z)
\,, \nn \\
P^{(0)}_{gg}(z) &= 2 C_A \,\theta(z) P_{gg}(z) + \beta_0 \,\delta(1-z)
\,, \quad
&P^{(0)}_{gq_i}(z) &= P^{(0)}_{g\bar q_i}(z) = 2C_F \,\theta(z) P_{gq}(z)
\,,\end{align}
with the standard color-stripped splitting functions,
\begin{alignat}{3} \label{eq:p_ij}
P_{qq}(z) &= \cL_0(1 - z)(1 + z^2) + \frac{3}{2}\delta(1-z)
\,, \qquad &
P_{qg}(z) &= \theta(1-z) \bigl[ 1-2z(1-z) \bigr]
\,, \nn \\
P_{gg}(z) &= 2\mathcal{L}_0(1-z) \frac{(1-z + z^2)^2}{z}
\,, \qquad &
P_{gq}(z) &= \theta(1-z) \frac{1+(1-z)^2}{z}
\,.\end{alignat}

\subsubsection{Calculation of beam function boundary terms}

The structure of \eq{modified_beam_function_master_formula} is exactly the same
as for the inclusive beam function $\cI_{ij}(t, z, \mu)$~\cite{Gaunt:2014xga, Gaunt:2014cfa},
except for the different boundary terms $I^{(n)}_{ij}(z) \neq \tI^{(n)}_{ij}(z)$.
The definition in \eq{def_modified_beam_function} implies for the matching coefficients
\begin{align} \label{eq:projection_matching_coefficient}
\tcI_{ij}(t, z, \mu) = \int \! \df^2 \vec{k}_T \, \cI_{ij}\Bigl(t - \frac{k_T^2}{2}, \vec{k}_T, z, \mu\Bigr)
\,,\end{align}
which we use to calculate $\tI^{(n)}_{ij}(z)$
to the extent that the double-differential matching coefficients are known,
i.e., to one loop for $i = g$~\cite{Jain:2011iu}
and two loops for $i = q$~\cite{Jain:2011iu, Gaunt:2014xxa}.
Note that for $i = g$, the integral over all $\vec{k}_T$  leaves behind only the
polarization-independent piece of the double-differential gluon beam function.
We have also verified that the $\mu$-dependent pieces $\propto \cL_n(t, \mu)$
obtained from \eq{projection_matching_coefficient} agree with the RGE prediction \eq{modified_beam_function_master_formula},
i.e., we have explicitly checked that the projection and the RGE commute.

At one loop, we find that the following simple relation holds for all partonic channels,
\begin{align} \label{eq:modified_beam_function_at_one_loop_is_simple}
\tcI^{(1)}_{ij} (t, z, \mu)
= \cI_{ij}^{(1)} (t, z, \mu) + \delta(t) \, P^{(0)}_{ij}(z)  \ln \frac{2z}{1+z}
\,,\end{align}
where $\cI_{ij}^{(1)}$ is the one-loop matching coefficient for the inclusive beam function.
Explicitly, the one-loop finite terms of the modified beam function are given by
\begin{align}
\tI^{(1)}_{q_iq_j} = \tI^{(1)}_{\bar{q}_i\bar{q}_j} \equiv \delta_{ij} \tI^{(1)}_{qq}(z)
&= 2C_F \, \delta_{ij} \, \theta(z) \Bigl[ \cL_1(1-z)(1+z^2) - \frac{\pi^2}{6} \delta(1-z) + \theta(1-z)(1 - z) + P_{qq}(z)\ln \frac{2}{1+z} \Bigr]
\,, \nn \\
\tI^{(1)}_{q_ig} = \tI^{(1)}_{\bar{q}_ig} = \tI^{(1)}_{qg}(z)
&= 2T_F \, \theta(z) \Bigl[ P_{qg}(z) \ln \frac{2(1-z)}{1+z} + \theta(1-z) \, 2z(1-z)\Bigr]
\,, \nn \\
\tI_{gg}^{(1)}(z)
&= 2C_A \, \theta(z) \Bigl[ \cL_1(1-z) \frac{2(1-z + z^2)^2}{z} - \frac{\pi^2}{6} \delta(1-z) + P_{gg}(z) \ln \frac{2}{1+z} \Bigr]
\,, \nn \\
\tI^{(1)}_{gq_i} = \tI^{(1)}_{g\bar{q}_j} = \tI_{gq}^{(1)}(z)
&= 2C_F \, \theta(z) \Bigl[ P_{gq}(z)\ln \frac{2(1-z)}{1+z} + \theta(1-z) \, z \Bigr]
\,.\end{align}
We decompose the two-loop quark finite terms $\tI_{ij}^{(2)}(z)$ by their flavor structure as
\begin{align} \label{eq:beam_flavor_decomposition}
\tI_{q_i q_j}^{(2)}(z)
= \tI_{\bq_i \bq_j}^{(2)}(z)
&= \delta_{ij} \tI_{qqV}^{(2)}(z) + \tI_{qqS}^{(2)}(z)
\,, \qquad
\tI_{q_i \bq_j}^{(2)}(z)
= \tI_{\bq_i q_j}^{(2)}(z)
= \delta_{ij} \tI_{q\bq V}^{(2)}(z) + \tI_{qqS}^{(2)}(z)
\,,\nn\\
\tI_{q_i g}^{(2)}(z)
= \tI_{\bq_i g}^{(2)}(z)
&= \tI_{q g}^{(2)}(z)
\,.\end{align}
As was done for $I^{(2)}_{ij}(z)$ in \refscite{Gaunt:2014xga, Gaunt:2014cfa},
we find it convenient to pull common rational factors
out of recurring terms with transcendental weight three ($\tilde{S}_3, \tilde{T}_3, \tilde{U}_3, \tilde{V}_3, \tilde{R}_{\dots}$),
and group terms of lower transcendental weight separately by color factor and flavor structure ($\tilde{C}_{\dots}$).
We also pull out a conventional factor of four:
\begin{align} \label{eq:two_loop_finite_terms}
\tilde{I}^{(2)}_{qqV}(z)
&= 4 C_F^2 \Bigl\{ D_{qqV,C_F}(z) - \frac{2}{1-z}\tilde{T}_3(z) + \frac{1+z^2}{1-z}\bigl[\tilde{V}_3(z) - 2\tilde{U}_3(z)\bigr] + \tilde{C}_{qqV,C_F}(z)\Bigr\}
\nn \\ & \quad
+ 4 C_F C_A \Bigl\{ D_{qqV,C_A}(z) + \frac{1+z^2}{1-z}\bigl[\tilde{U}_3(z)+\tilde{R}_{qqV}(z)\bigr] + \tilde{C}_{qqV,C_A}(z)\Bigr\}
+  4 C_F \beta_0 \Bigl[D_{qqV,\beta_0}(z) + \tilde{C}_{qqV,\beta_0}(z)\Bigr]
\,, \nn \\
\tilde{I}^{(2)}_{q\bar{q}V}(z)
&= 4 C_F (2C_F - C_A) \Bigl[\frac{1+z^2}{1+z}\tilde{S}_3(z) + \tilde{C}_{q\bar{q}V}(z) \Bigr]
\,, \nn \\
\tilde{I}^{(2)}_{qqS}(z)
&=  4 C_F T_F\Bigl[-2(1+z) \, \tilde{T}_3(z) + \tilde{C}_{qqS}(z) \Bigr]
\,, \nn \\
\tilde{I}^{(2)}_{qg}(z) &= 4 T_F C_F\Bigl\{-2(1-z)^2 \, \tilde{T}_3(z) + P_{qg}(z)\bigl[\tilde{V}_3(z)+\tilde{R}_{qg,C_F}(z)\bigr] + \tilde{C}_{qg,C_F}(z) \Bigr\}
\nn \\ & \quad
+ 4 T_F C_A\Bigl\{ -2(1+4z) \, \tilde{T}_3(z) - P_{qg}(z)\bigl[\tilde{U}_3(z)+\tilde{R}_{qg,C_A}(z)\bigr] + P_{qg}(-z) \, \tilde{S}_3(z)
+ \tilde{C}_{qg,C_A}(z) \Bigr\}
\,.\end{align}
Here, overall factors of $\theta(z) \, \theta(1-z)$ are understood, but omitted for brevity.
The $D_{qqV, \dots}(z)$ contain all distributional terms in $1-z$
and are the same as for the standard inclusive beam function~\cite{Gaunt:2014xga},
\begin{align}
D_{qqV,C_F}(z) &= (1+z^2)\Bigl[\cL_3(1-z) - \frac{5\pi^2}{6}\cL_1(1-z) + 4\zeta_3\cL_0(1-z)\Bigr] + \frac{7\pi^4}{120}\delta(1-z)
\,, \nn\\
D_{qqV,C_A}(z) &= (1+z^2)\Bigl[\Bigl(\frac{2}{3} - \frac{\pi^2}{6}\Bigr)\cL_1(1-z) + \Bigl(-\frac{8}{9} + \frac{7\zeta_3}{2}\Bigr)\cL_0(1-z)\Bigr] + \Bigl(\frac{52}{27} - \frac{\pi^2}{6} - \frac{\pi^4}{36}\Bigr)\delta(1-z)
\,, \nn\\
D_{qqV,\beta_0}(z) &= (1+z^2)\Bigl[-\frac{1}{4}\cL_2(1-z) + \frac{5}{6}\cL_1(1-z) + \Bigl(-\frac{7}{9} + \frac{\pi^2}{12}\Bigr)\cL_0(1-z)\Bigr] + \Bigl(\frac{41}{27} - \frac{5\pi^2}{24} - \frac{5\zeta_3}{6}\Bigr)\delta(1-z)
\,.\end{align}
All remaining terms in \eq{two_loop_finite_terms} are integrable for $z\to 1$.
Their full expressions are lengthy and are available from the authors upon request.
As an example of the structures that occur, we give
\begin{align}
\tilde{S}_3(z) &= 2 G\bigl(-1, -1, -\tfrac{1}{2}; z\bigr)
- 3 G\bigl(-1, 0, -\tfrac{1}{2}; z\bigr)
- 3 G\bigl(0, -1, -\tfrac{1}{2}; z\bigr)
+ 4 G\bigl(0, 0, -\tfrac{1}{2}; z\bigr)
\nn \\[0.4em] & \quad
- 2 H(-1, -1, 0; z)
+ 2 H(-1, 0, -1; z)
- 3 H(-1, 0, 0; z)
- 2 H(-1, 1, 0; z)
+ H(0, -1, 0; z)
\nn \\ & \quad
- H(0, 0, -1; z)
- H(0, 0, 1; z)
- 2 H(1, -1, 0; z)
- \frac{\ln^3(z)}{12} - 2H(-1,0;z)\ln(1-z)
\nn \\ & \quad + \ln(2) \Bigl[-\frac{\pi^2}{6} + G\bigl(-1, -\tfrac{1}{2}; z\bigr) - G\bigl(0, -\tfrac{1}{2}; z\bigr) - 2 H(-1, 0; z) + \frac{\ln^2(z)}{2}\Bigr]
\nn \\ & \quad
+ \ln(1+z)\Bigl[-\frac{5\pi^2}{12} - G\bigl(-1, -\tfrac{1}{2}; z\bigr) + G\bigl(0, -\tfrac{1}{2}; z\bigr) \Bigr]
\nn \\ & \quad
+ \ln(z) \Bigl[\frac{\pi^2}{4} + 2 G\bigl(-1, -\tfrac{1}{2}; z\bigr) - 2 G\bigl(0, -\tfrac{1}{2}; z\bigr) + 2 H(-1,0,z)\Bigr]
+ \frac{9 \zeta_3}{4}
\,.\end{align}
Here, we have used the recent \texttt{PolyLogTools} package~\cite{Duhr:2019tlz}
to convert all polylogarithms of rational functions of $z$
to standard harmonic polylogarithms $H(a_1, \dots, a_n; z)$ as well as
multiple polylogarithms $G(a_1, \dots, a_n; z)$ of $z \in [0,1]$.
The latter are as defined in \refcite{Duhr:2019tlz},
and for all $a_i = 0,\pm 1$ reduce to standard harmonic polylogarithms up to a sign.
We find no evidence for a simple generalization of the one-loop relation
\eq{modified_beam_function_at_one_loop_is_simple} at two loops.
Finally, we note that the two-loop $\tI_{ij}^{(2)}(z)$ for the modified beam function
are substantially more complicated than those for the standard inclusive beam function.
For example, the latter does not involve polylogarithms with fractional
weights, which only arise from the projection integral in \eq{projection_matching_coefficient}.
We expect that similarly the three-loop cross section in terms of $(q^+, q^-, \vec{q}_T)$
will have a much simpler structure than in terms of $(Q, Y, \vec{q}_T)$.

\subsection{Breakdown of NNLO validation into partonic channels}

Here, we provide the breakdown of the numerical validation of
\eq{factorization_collinear_endpoint_Q_Y} at NNLO in
\fig{sing_nons_gamma_nnlo_all_xb_1em2} into individual partonic channels.
Following \texttt{Vrap}~\cite{Anastasiou:2003ds}, we take the $ij = q\bar{q}$
channel to include all topologies where $i$ and $j$ are part of the same quark
line. The leading-power limit of these diagrams corresponds to the $qqV$ beam
function matching coefficient in the decomposition in  \eq{beam_flavor_decomposition}.
In addition, the $q\bar{q}$ channel also includes purely nonsingular
contributions with topologies $q\bar{q} \to g \to q\bar q V$. We then take the
$qq'$ channel to include the remaining quark-initiated processes, which at
leading power reduces to the sum of the $qqS$ and $q\bar{q}V$ beam function
contributions  in \eq{beam_flavor_decomposition}. The $qg$ channel maps onto the
$qg$ beam function contribution at leading power, while the $gq$ and $gg$
channels are purely nonsingular.

\begin{figure*}[t]
\includegraphics[width=0.48\textwidth]{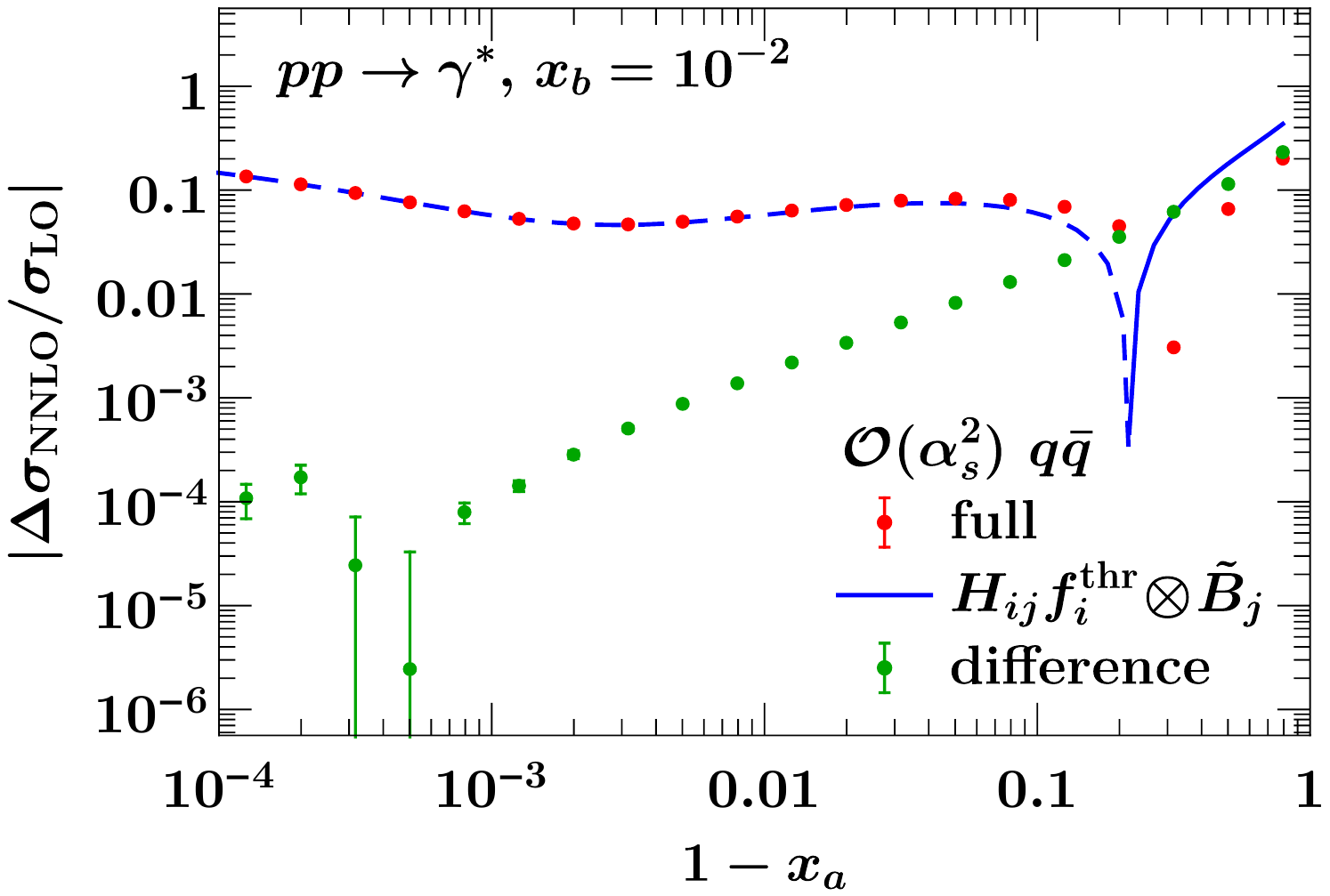}%
\hfill%
\includegraphics[width=0.48\textwidth]{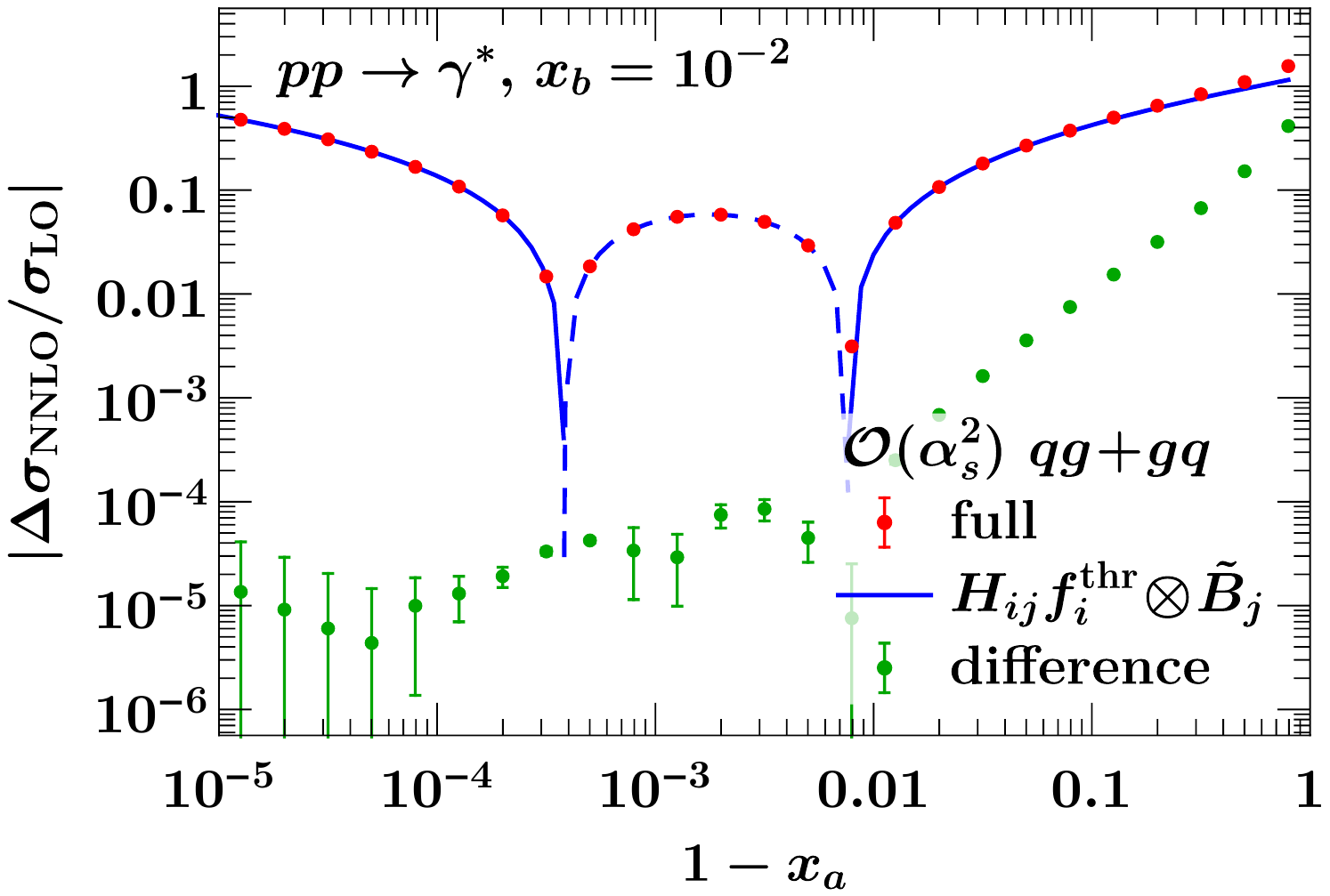}%
\\[1ex]
\includegraphics[width=0.48\textwidth]{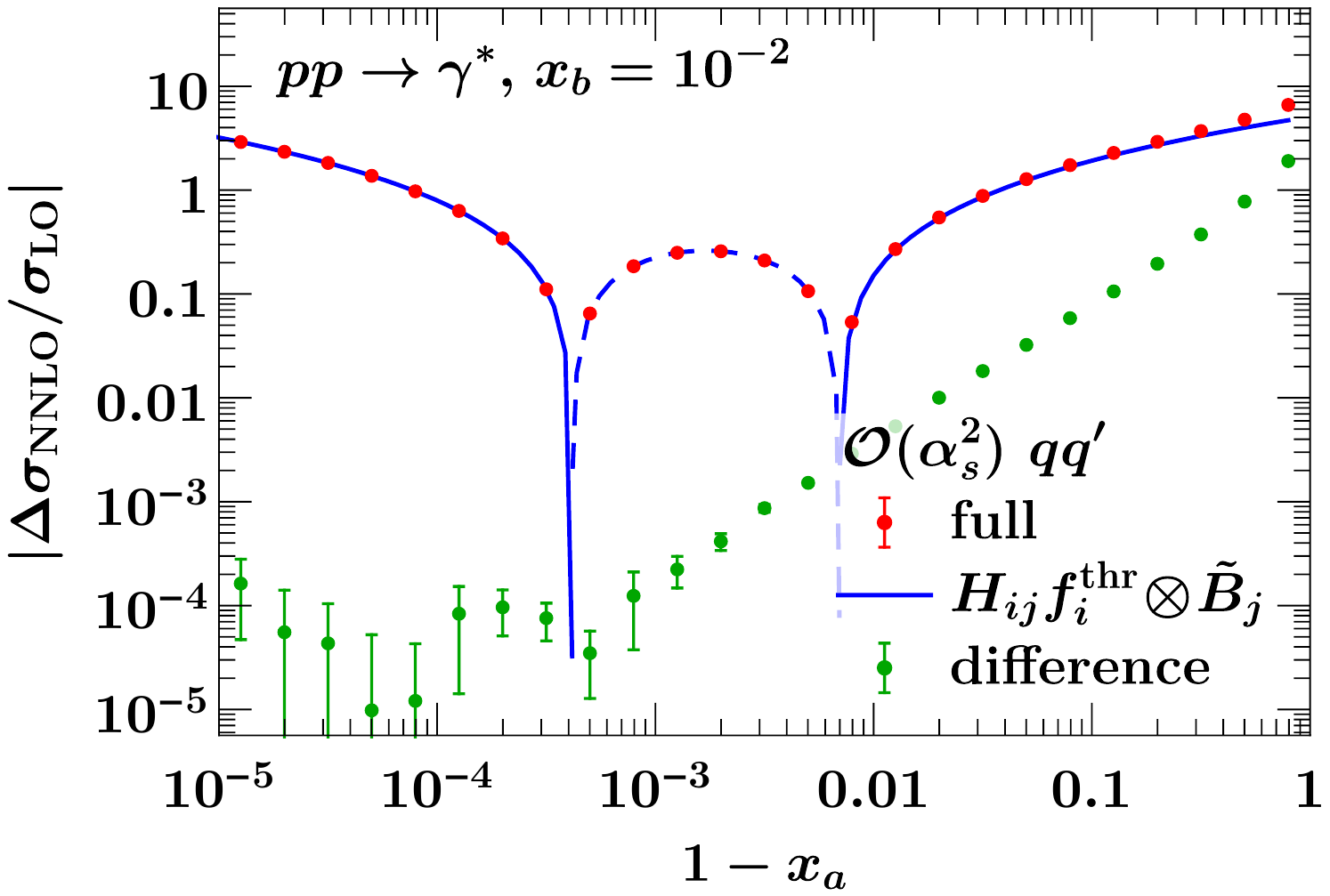}%
\hfill%
\includegraphics[width=0.48\textwidth]{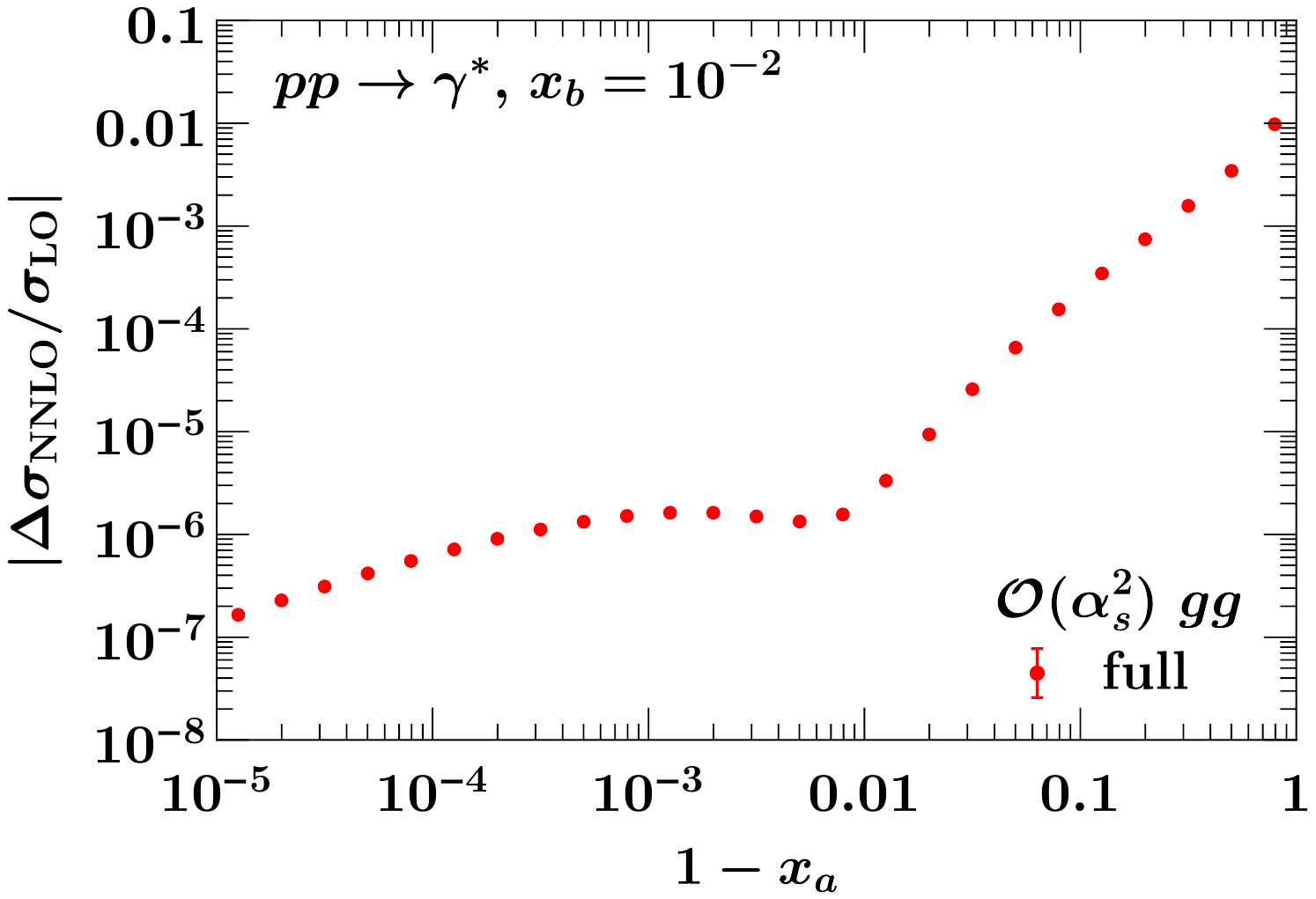}%
\caption{Breakdown of \fig{sing_nons_gamma_nnlo_all_xb_1em2} into partonic channels.
Shown are the $\ord{\as^2}$ contribution to $\df\sigma/\df x_a\df x_b$
predicted for $x_a \to 1$ by \eq{factorization_collinear_endpoint_Q_Y} (blue),
the full result from \texttt{Vrap} (red). In all cases, their difference (green)
vanishes like a power as $1 - x_a \to 0$, as it must.
The $gg$ channel is power suppressed, so its full result by itself vanishes like
a power. The error bars indicate the integration uncertainties.
Dashing in the blue line indicates a negative result.
\label{fig:sing_nons_gamma_nnlo_channels_xb_1em2}}
\end{figure*}

The results are shown in \fig{sing_nons_gamma_nnlo_channels_xb_1em2}.
In all cases, the prediction of \eq{factorization_collinear_endpoint_Q_Y} is in excellent
agreement with the singular limit of the full calculation,
with their difference vanishing as a power of $1-x_a$ as it should.
The excellent numerical stability of \texttt{Vrap} for the nondiagonal channels
allows us to extend the check down to $1-x_a = 10^{-5}$,
where it becomes limited by MC statistics.
For the $q\bar{q}$ channel, we start to see a systematic deviation at the $10^{-4}$ level
below $1-x_a \lesssim 10^{-4}$. We observe a similar deviation already at NLO,
where the partonic cross sections agree analytically. We thus attribute this to a
systematic effect in the PDF integrations in \texttt{Vrap}.

\subsection{\boldmath Results for Drell-Yan at $\mu = Q/2$ and for gluon-fusion Higgs production}

\begin{figure*}[t]
\includegraphics[width=0.48\textwidth]{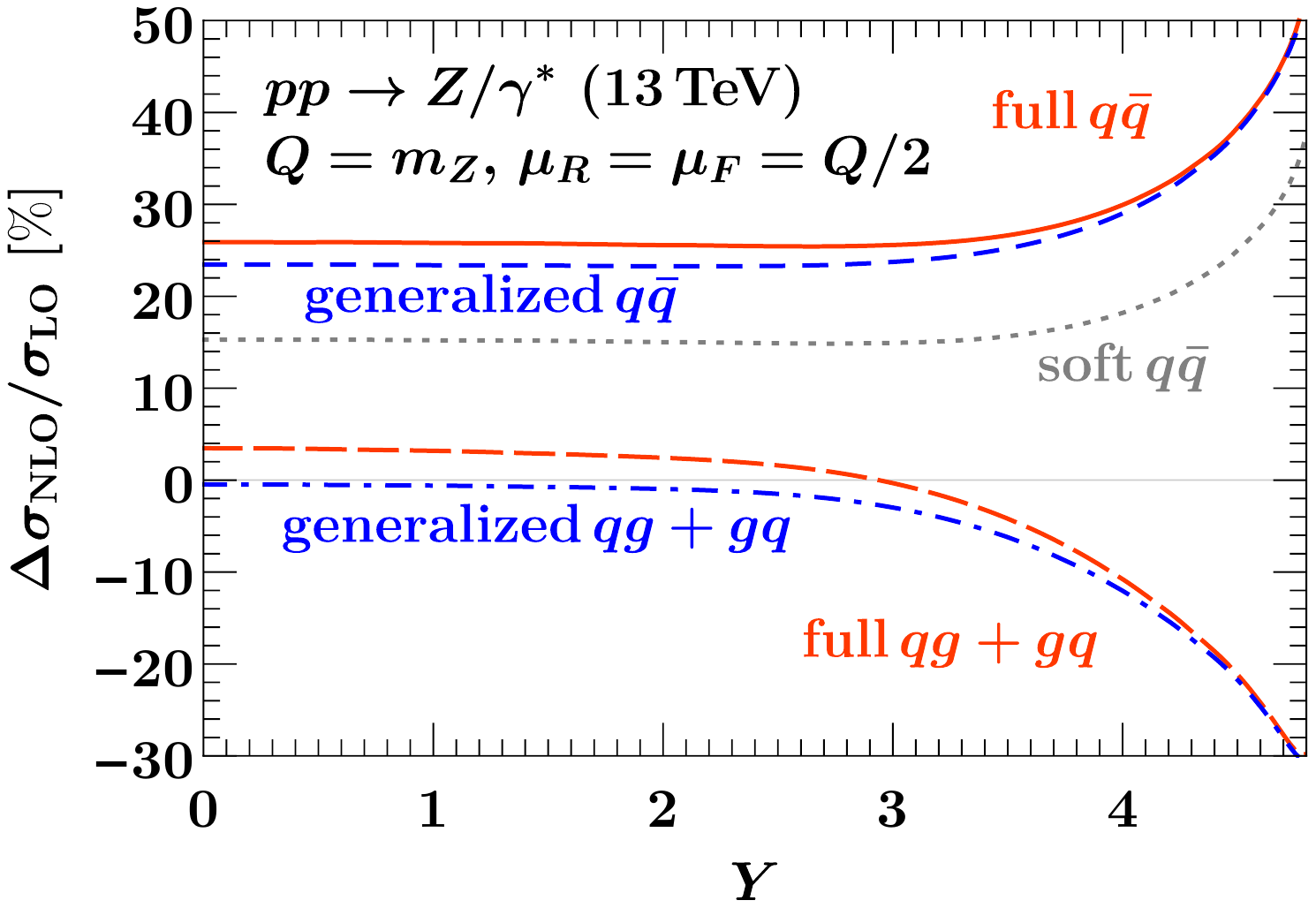}%
\hfill%
\includegraphics[width=0.48\textwidth]{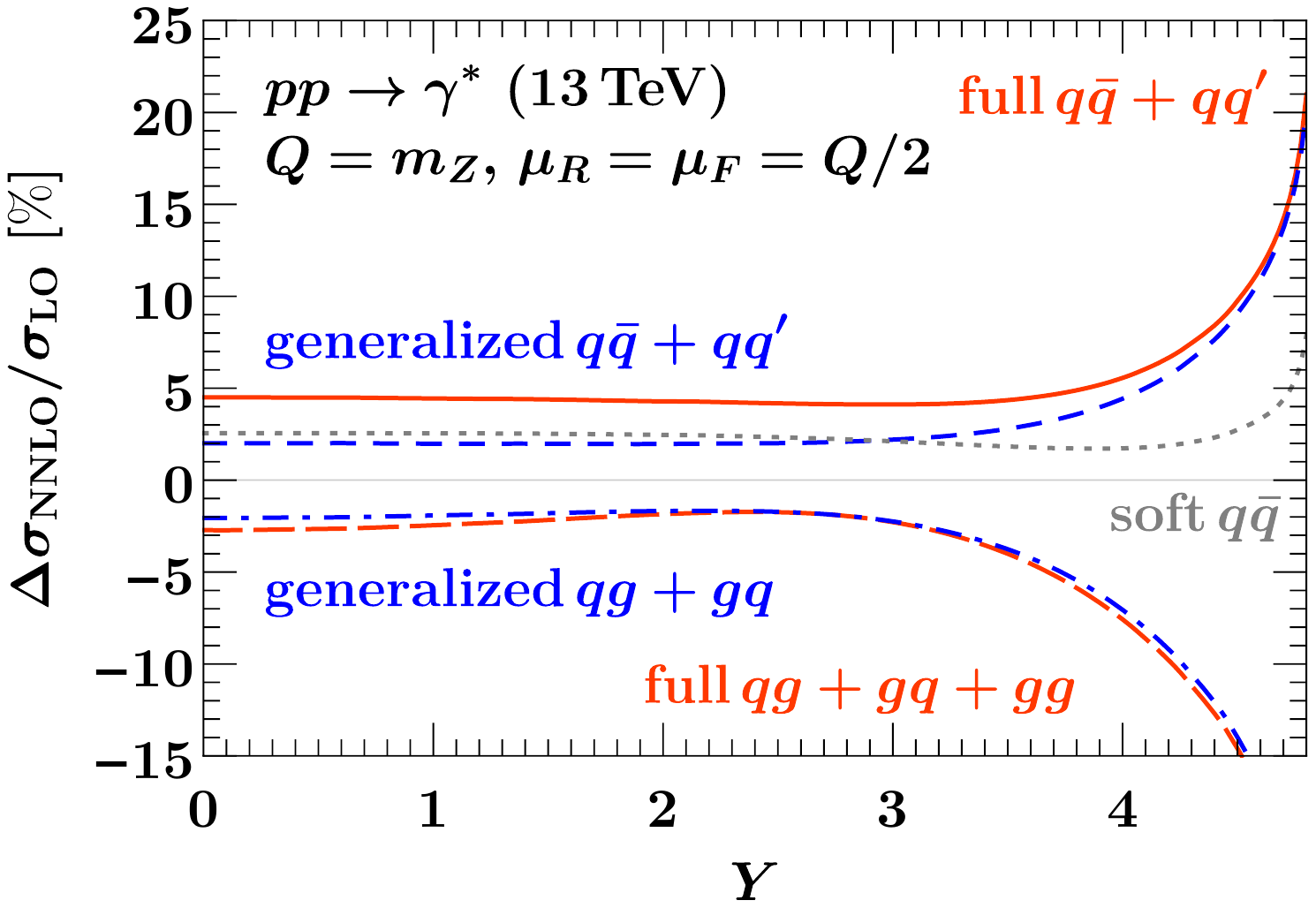}%
\\[1ex]
\includegraphics[width=0.48\textwidth]{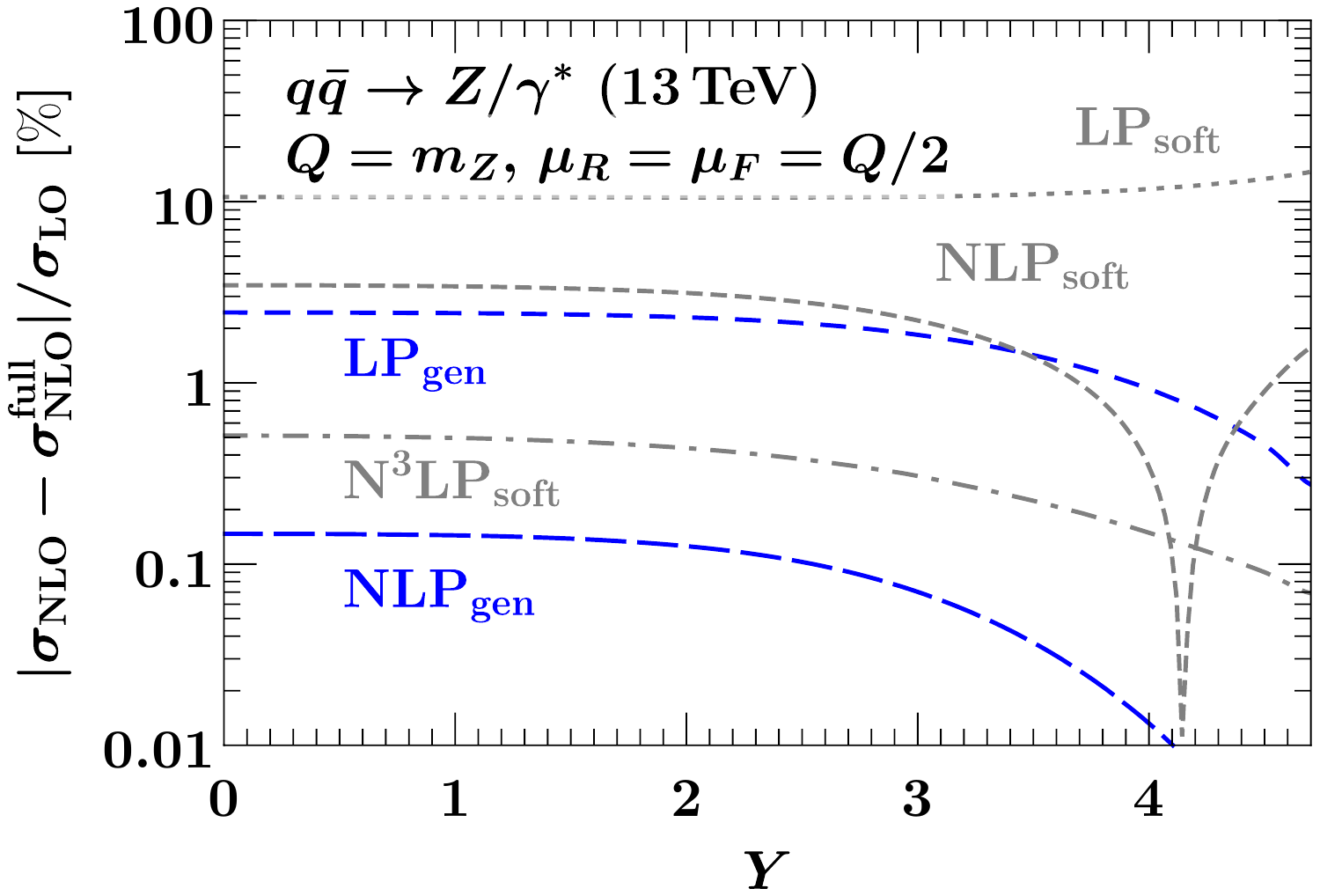}%
\hfill%
\includegraphics[width=0.48\textwidth]{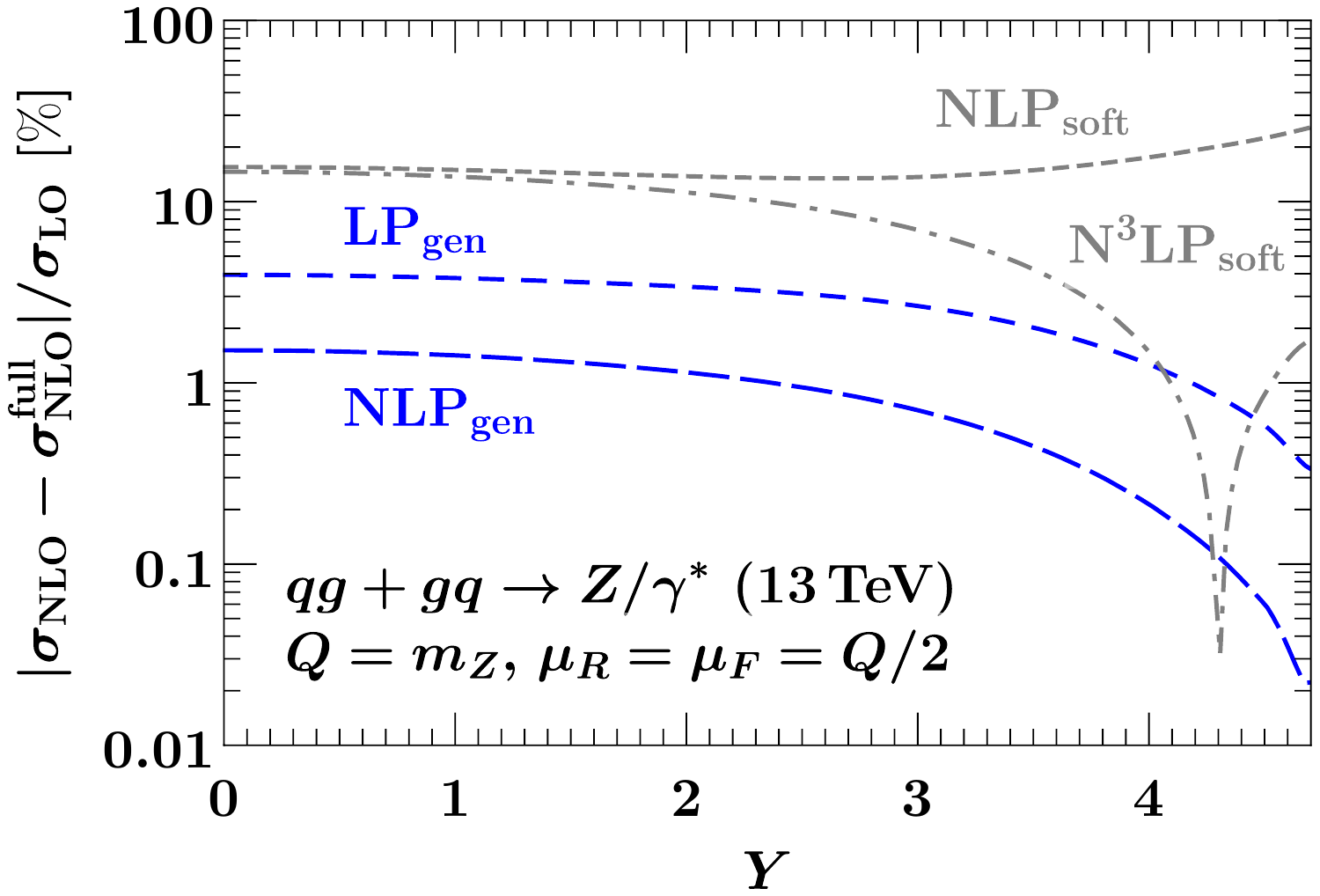}%
\caption{Top row: Generalized threshold approximation of the $\ord{\as}$ (top left) and $\ord{\as^2}$ contribution (top right) to the Drell-Yan rapidity spectrum $\sigma \equiv \df \sigma/(\df Q \df Y)$
at $\mu = Q/2$ normalized to the LO result.
This is the analog of \fig{generalized_threshold_approximation_drell_yan}.
\\
Bottom row: Convergence of the generalized and soft threshold expansions for the $q\bq$ (bottom left) and $qg+gq$ channels (bottom right) for $\mu = Q/2$.
This is the analog of \fig{generalized_threshold_approximation_drell_yan_convergence}.
\label{fig:generalized_threshold_approximation_drell_yan_muFO_Q_2}}
\end{figure*}

\begin{figure*}[t]
\centering
\includegraphics[width=0.48\textwidth]{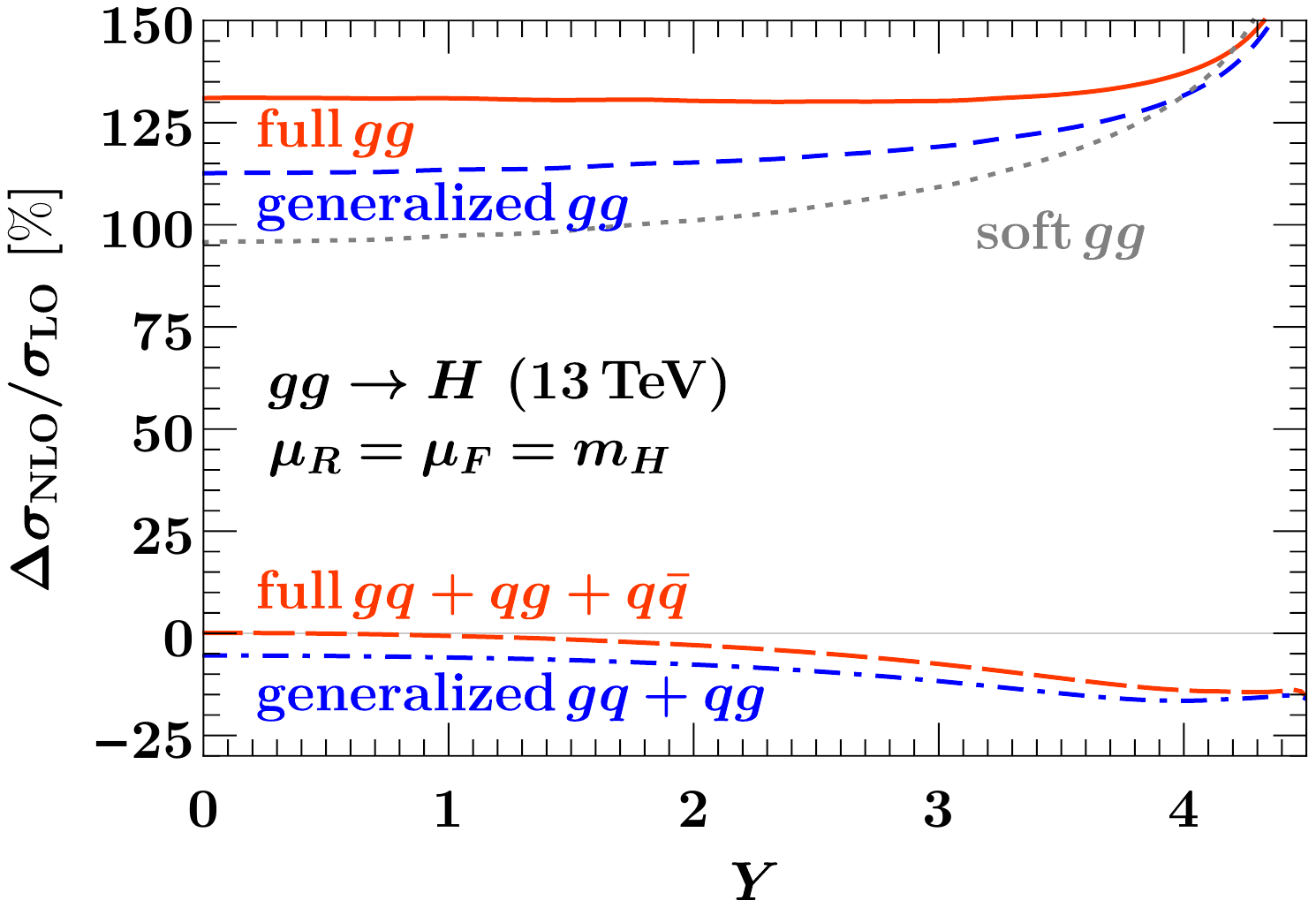}%
\hfill%
\includegraphics[width=0.48\textwidth]{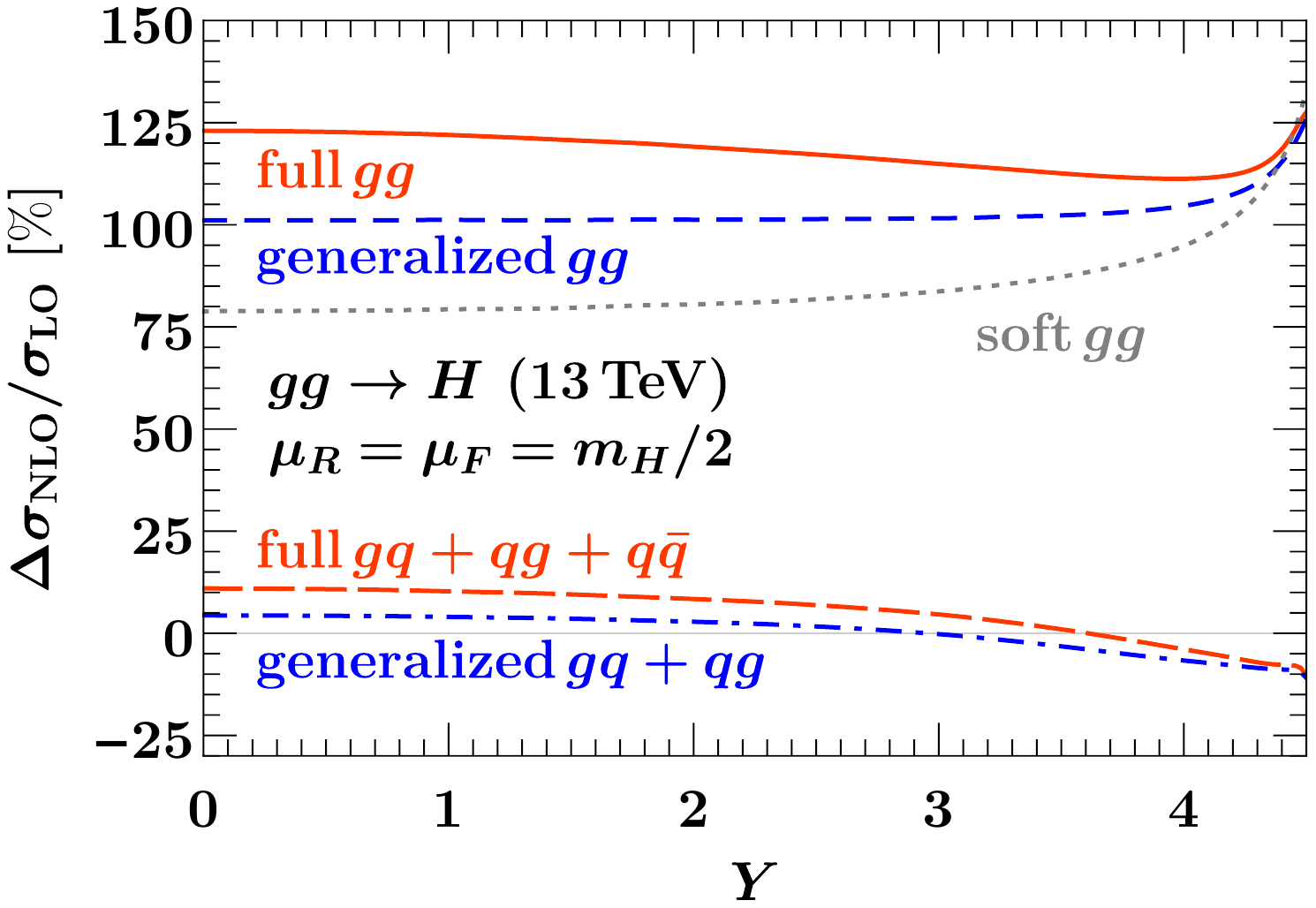}%
\\[1ex]
\includegraphics[width=0.48\textwidth]{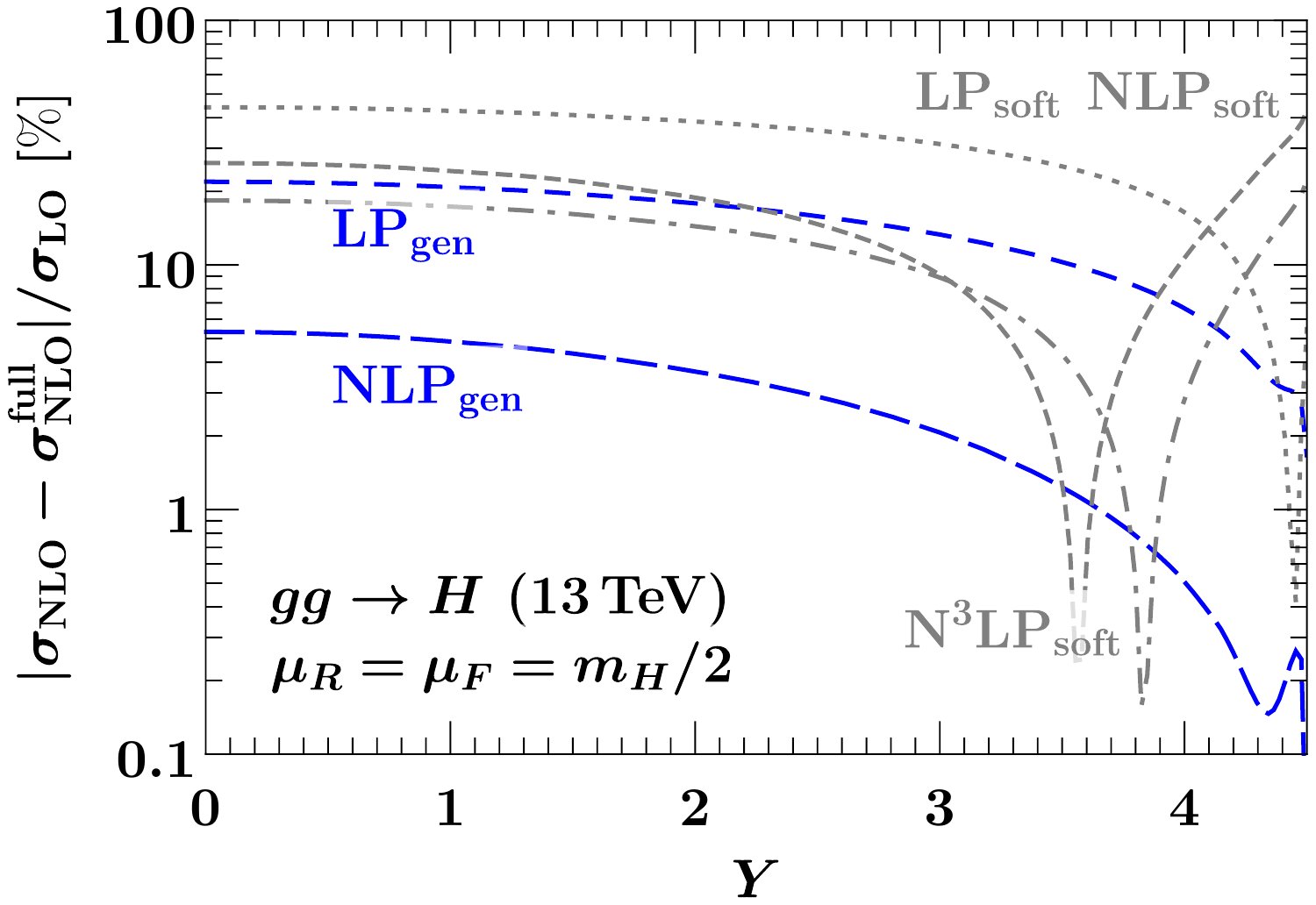}%
\hfill%
\includegraphics[width=0.48\textwidth]{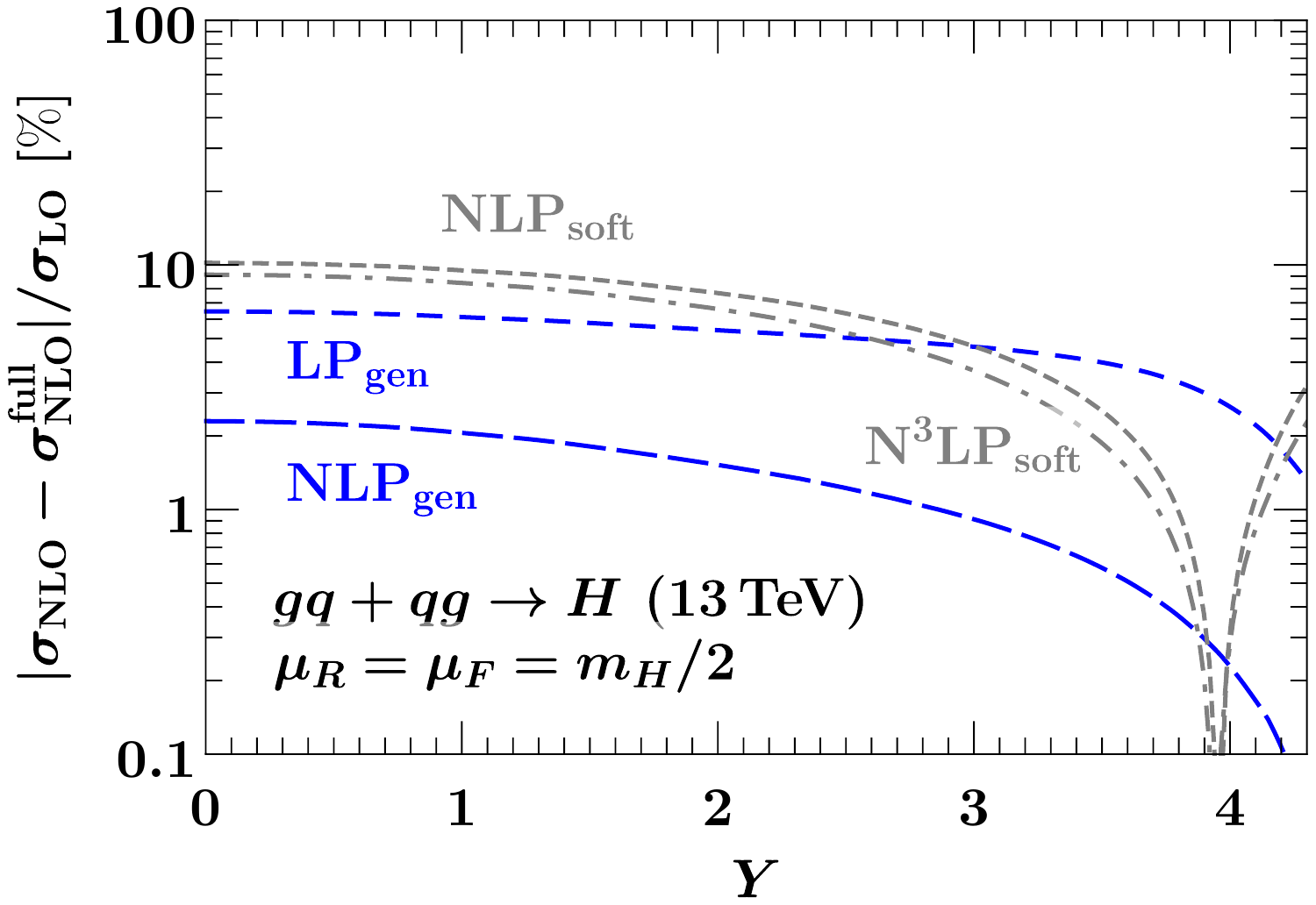}%
\caption{Top row: Generalized threshold approximation of the $\ord{\as}$
contribution to the $gg\to H$ rapidity spectrum $\sigma \equiv \df \sigma/\df Y$
at $\mu = m_H$ (top left) and $\mu = m_H/2$ (top right) normalized to the LO result.
This is the analog of \fig{generalized_threshold_approximation_drell_yan}.
\\
Bottom row: Convergence of the generalized and soft threshold expansions for the $gg$
(bottom left) and $gq+qg$ channels (bottom right) for $\mu = m_H/2$.
This is the analog of \fig{generalized_threshold_approximation_drell_yan_convergence}.
\label{fig:generalized_threshold_approximation_ggH}}
\end{figure*}

Here, we provide additional results for the generalized threshold expansion at
fixed order. The analogs of \figs{generalized_threshold_approximation_drell_yan}
{generalized_threshold_approximation_drell_yan_convergence} for the Drell-Yan
rapidity spectrum at a different scale choice $\mu = Q/2$ are shown in
\fig{generalized_threshold_approximation_drell_yan_muFO_Q_2}, and for the $gg\to
H$ NLO rapidity spectrum at $\mu = m_H$ and $\mu = m_H/2$ in
\fig{generalized_threshold_approximation_ggH}.

When performing any threshold expansion for different scale choices, there are
several options how to treat the terms in the partonic cross section that are
predicted by the running of the PDFs or $\alpha_s$. One option is to expand
these terms to the working order in the threshold expansion. This ensures that
the partonic cross section has homogeneous power counting at any scale, but
leaves the running of the PDFs and the coupling uncanceled beyond the working
order. Another option is to threshold-expand the partonic cross section at a
given reference scale and treat the running exactly. This leads to a privileged
scale where the expansion was performed, but ensures the cancellation of $\as$
and PDF running to all powers (up to higher orders in $\as$). In the following
results, we choose the first option for definiteness. The difference between the
two approaches could serve as a way to estimate the size of power corrections.

For Drell-Yan at $\mu = Q/2$, the generalized threshold expansion performs
similarly well as for $\mu = Q$ in the main text, and again much better than the
soft expansion. For $gg\to H$, the generalized threshold expansion again
performs in a manner clearly superior to the soft one. Here, the increment from the
leading-power soft to the leading-power generalized approximation of the $gg$
channel at NLO is roughly comparable to the piece still missing to the full
result; either contribution amounts to $\ord{20\%}$ in units of the Born cross
section. This is consistent with the expectation that for a gluon-induced
process, hard central radiation plays a larger role than for Drell-Yan. The
shape of the NLO contributions at large $Y$ is well captured by the
leading-power generalized approximation for both the $gg$ and $gq+qg$ channel.
The leading-power soft approximation for the $gg$ channel (on close inspection)
turns out to be off at large $Y$, and in both channels there is barely any
convergence beyond leading power in the soft expansion at any $Y$.

\end{document}